\documentclass[fleqn,10pt]{wlscirep}
\title{Asymmetric Biphoton Generation under Ground-State Decoherence and Phase Mismatch in a Cold Atomic Ensemble}

\author[1,2]{Jiun-Shiuan Shiu}
\author[1]{Chang-Wei Lin}
\author[1,2*]{Yong-Fan Chen}
\affil[1]{Department of Physics, National Cheng Kung University, Tainan 70101, Taiwan}
\affil[2]{Center for Quantum Frontiers of Research \& Technology, Tainan 70101, Taiwan}
\affil[*]{yfchen@mail.ncku.edu.tw}

\begin{abstract}

We present an experimental investigation of how ground-state decoherence and phase mismatch influence biphoton generation in double-$\Lambda$ spontaneous four-wave mixing (SFWM) within a cold atomic ensemble. Our results reveal significant asymmetry in the Stokes and anti-Stokes photon generation rates, arising from the distinct effects of phase mismatch and ground-state decoherence. While phase mismatch primarily drives this asymmetry under minimal decoherence, larger decoherence further amplifies it, underscoring the complex interplay between these factors. Using the coincidence count rate representation, we provide insights into pairing ratios and demonstrate that the stimulated four-wave mixing process inherent in SFWM explains the observed phenomena. Interestingly, although ground-state decoherence reduces the generation of temporally correlated photons, it paradoxically enhances biphoton purity, as confirmed through conditional autocorrelation measurements. This counterintuitive phenomenon is reported here for the first time. Furthermore, unconditional autocorrelation measurements show that the generated photons follow a thermal-state distribution, consistent with theoretical predictions. This study advances the understanding of biphoton generation dynamics and temporal photon correlations in SFWM, offering valuable insights for optimizing SFWM-based biphoton sources and their applications in quantum technologies.

\end{abstract}

\begin{document}

\flushbottom
\maketitle
\thispagestyle{empty}


\newcommand{\figone}
{
  \begin{figure}[t]
  \centering
  \includegraphics[width = 16 cm]{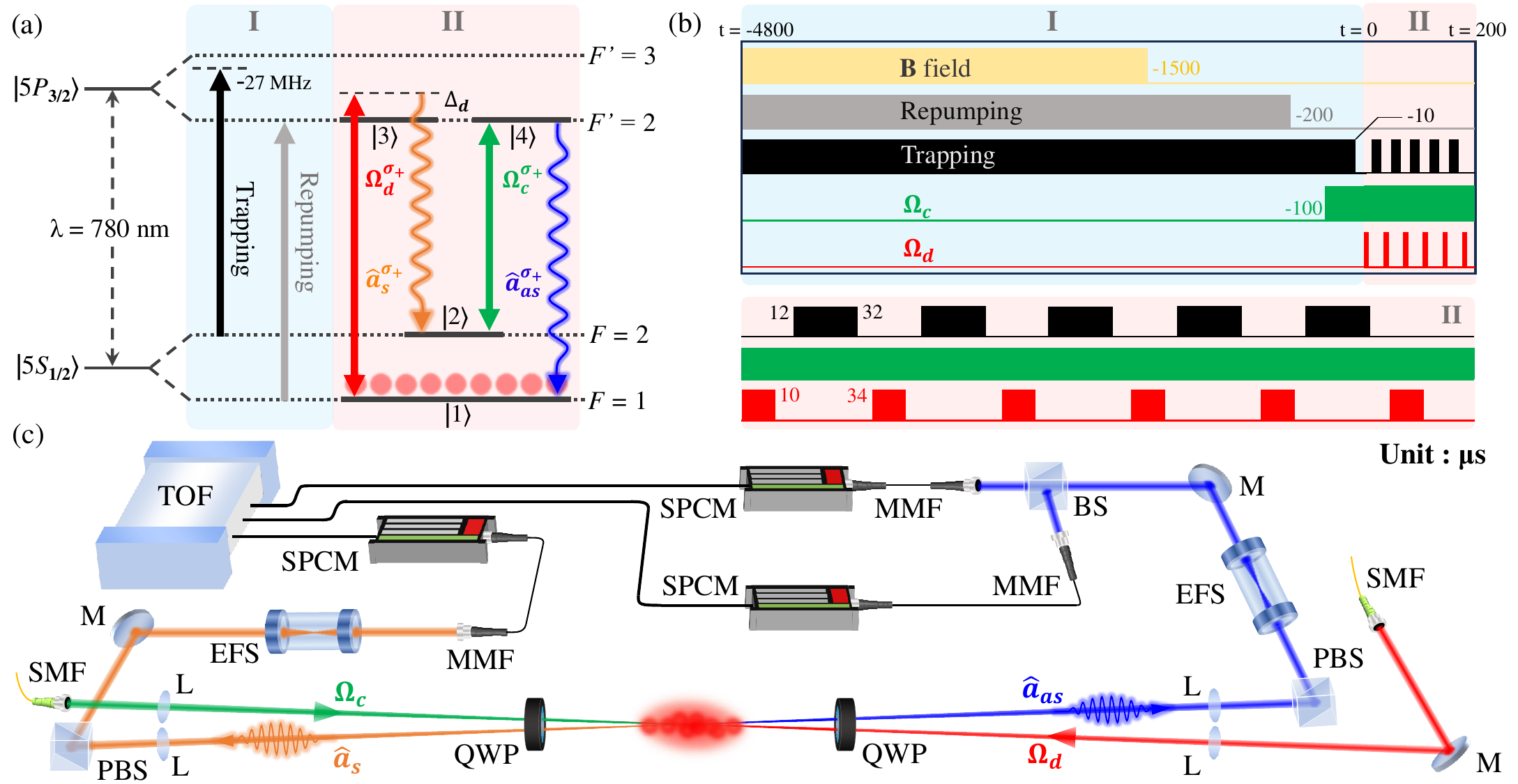}
  \caption
   {
     Schematic of the biphoton generation scheme via double-$\Lambda$ SFWM. (a) Energy-level diagram of $^{87}$Rb relevant to the SFWM experiment. Multiple Zeeman sub-levels are omitted for clarity. The diagram is divided into two regions: Region I is used for cooling $^{87}$Rb atoms, while Region II is dedicated to biphoton generation. In this experiment, $|3\rangle$ and $|4\rangle$ denote the same physical energy level. (b) Timing sequence of the experiment. Each 5-ms cycle comprises $4800 \,\,\mathrm{\mu s}$ for atomic cooling (Region I) and $200 \,\,\mathrm{\mu s}$ for biphoton generation (Region II). The time origin (t = 0) is defined at the onset of the first driving pulse ($\Omega_d$), with six driving pulses applied per cycle. During intervals between pulses, both the trapping and coupling fields ($\Omega_c$) are switched on to reinitialize the atomic population in the ground state $|1\rangle$. (c) Experimental setup for the double-$\Lambda$ SFWM system. Key optical components are labeled: M, mirror; L, lens; BS, beam splitter; PBS, polarizing beam splitter; QWP, quarter-wave plate; SMF, single-mode fiber; MMF, multi-mode fiber; EFS, etalon filter set; SPCM, single-photon counting module; TOF, time-of-flight multiscaler.
   }
  \label{fig1}
  \end{figure}
}
  \newcommand{\figtwo} 
{
  \begin{figure}[t]
  \centering
  \includegraphics[width = 12 cm]{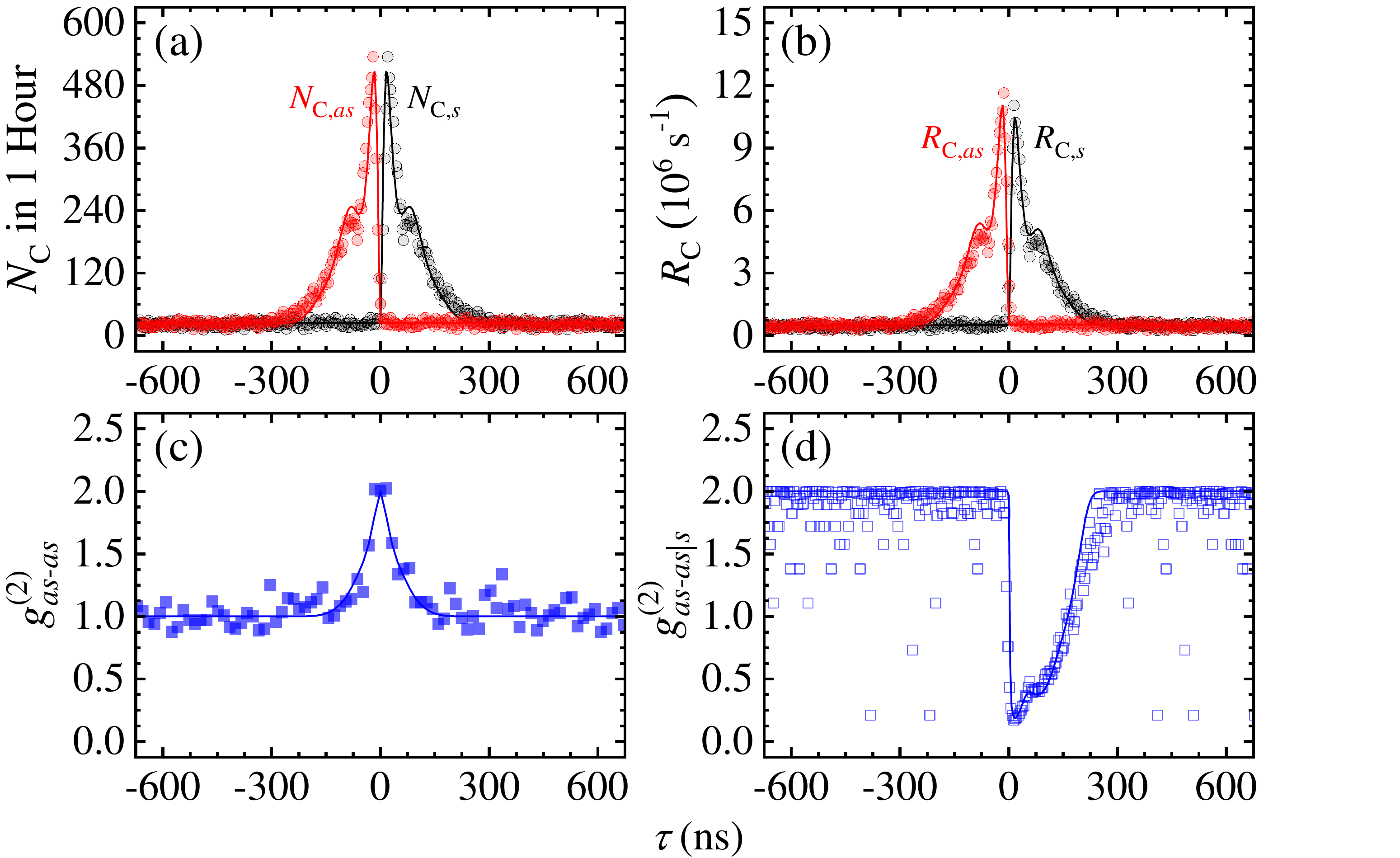}
  \caption
   {
     Asymmetric biphoton generation due to phase mismatch ($\Delta kL=0.37\pi$). Experimental parameters: ${\rm OD}=20$, $\Omega_d=\Omega_c=2\Gamma$, $\Delta_d=20\Gamma$, $\gamma_{21}=0.001\Gamma$, and time bin $\Delta\tau = 4 \,\, \mathrm{ns}$ for (a), (b), and (d), and $16 \,\, \mathrm{ns}$ for (c). (a) Black solid circles show the experimental coincidence counts $N_{{\rm C},s}$ with the Stokes channel as the trigger, while red solid circles show $N_{{\rm C},as}$ with the anti-Stokes channel as the trigger. (b) Experimental coincidence count rates $R_{{\rm C},s}^{\rm exp}$ and $R_{{\rm C},as}^{\rm exp}$ (black and red circles) derived from $N_{{\rm C},s}$ and $N_{{\rm C},as}$, respectively. Theoretical predictions $R_{{\rm C},s}^{\rm theo}$ and $R_{{\rm C},as}^{\rm theo}$ (black and red curves) are obtained from the theoretical model. In this case, the photon generation rates are $R_{s}=4.8 \times 10^5 \,\, \mathrm{s}^{-1}$ and $R_{as}=4.5 \times 10^5 \,\, \mathrm{s}^{-1}$. (c) Unconditional autocorrelation function $g_{as\text{-}as}^{(2)}$. Experimental data are shown as blue solid squares, and the theoretical curve is plotted as a blue line. (d) Conditional autocorrelation function $g_{as\text{-}as|s}^{(2)}$. Blue squares and the curve represent the experimental data and theoretical prediction.
   }
  \label{fig2}
  \end{figure}
}
\newcommand{\figthree} 
{
  \begin{figure}[t]
  \centering
  \includegraphics[width = 12 cm]{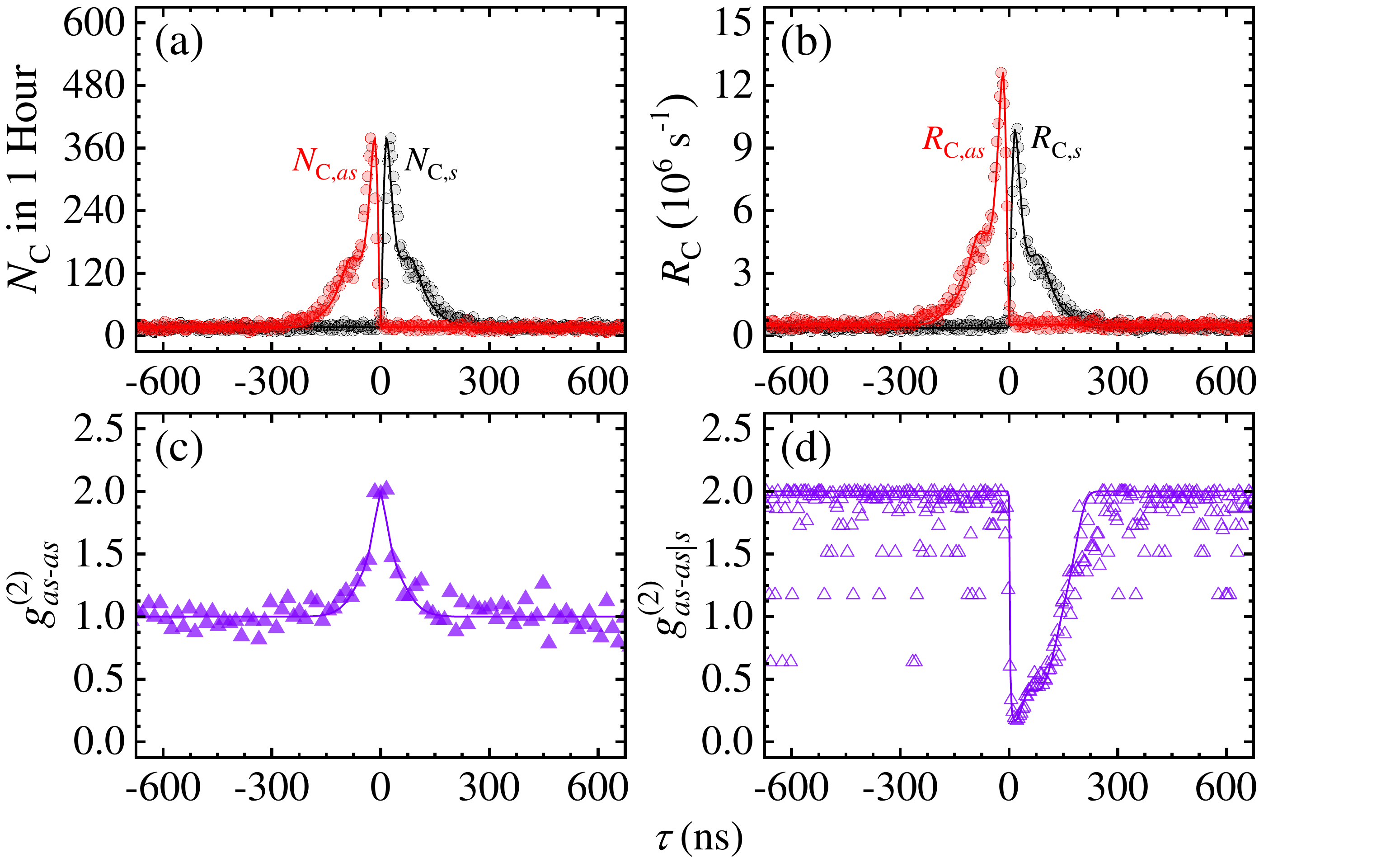}
  \caption
   {
     Asymmetric biphoton generation due to ground-state decoherence ($\gamma_{21}=0.1\Gamma$). Experimental parameters: ${\rm OD}=20$, $\Omega_d=\Omega_c=2\Gamma$, $\Delta_d=20\Gamma$, $\Delta kL=0.37\pi$, and time bin $\Delta\tau = 4 \,\, \mathrm{ns}$ for (a), (b), and (d), and $16 \,\, \mathrm{ns}$ for (c). (a) Black solid circles show the experimental coincidence counts $N_{{\rm C},s}$ with the Stokes channel as the trigger, while red solid circles show $N_{{\rm C},as}$ with the anti-Stokes channel as the trigger. (b) Experimental coincidence count rates $R_{{\rm C},s}^{\rm exp}$ and $R_{{\rm C},as}^{\rm exp}$ (black and red circles) derived from $N_{{\rm C},s}$ and $N_{{\rm C},as}$, respectively. Theoretical predictions $R_{{\rm C},s}^{\rm theo}$ and $R_{{\rm C},as}^{\rm theo}$ (black and red curves) obtained from the theoretical model. In this case, the photon generation rates are $R_{s}=4.8 \times 10^5 \,\, \mathrm{s}^{-1}$ and $R_{as}=3.8 \times 10^5 \,\, \mathrm{s}^{-1}$. (c) Unconditional autocorrelation function $g_{as\text{-}as}^{(2)}$. Experimental data are shown as purple solid triangles, and the theoretical curve is plotted as a purple line. (d) Conditional autocorrelation function $g_{as\text{-}as|s}^{(2)}$. Purple triangles and the curve represent the experimental data and theoretical prediction.
   }
  \label{fig3}
  \end{figure}
}

  \newcommand{\figfour} 
{
  \begin{figure}[t]
  \centering
  \includegraphics[width = 12 cm]{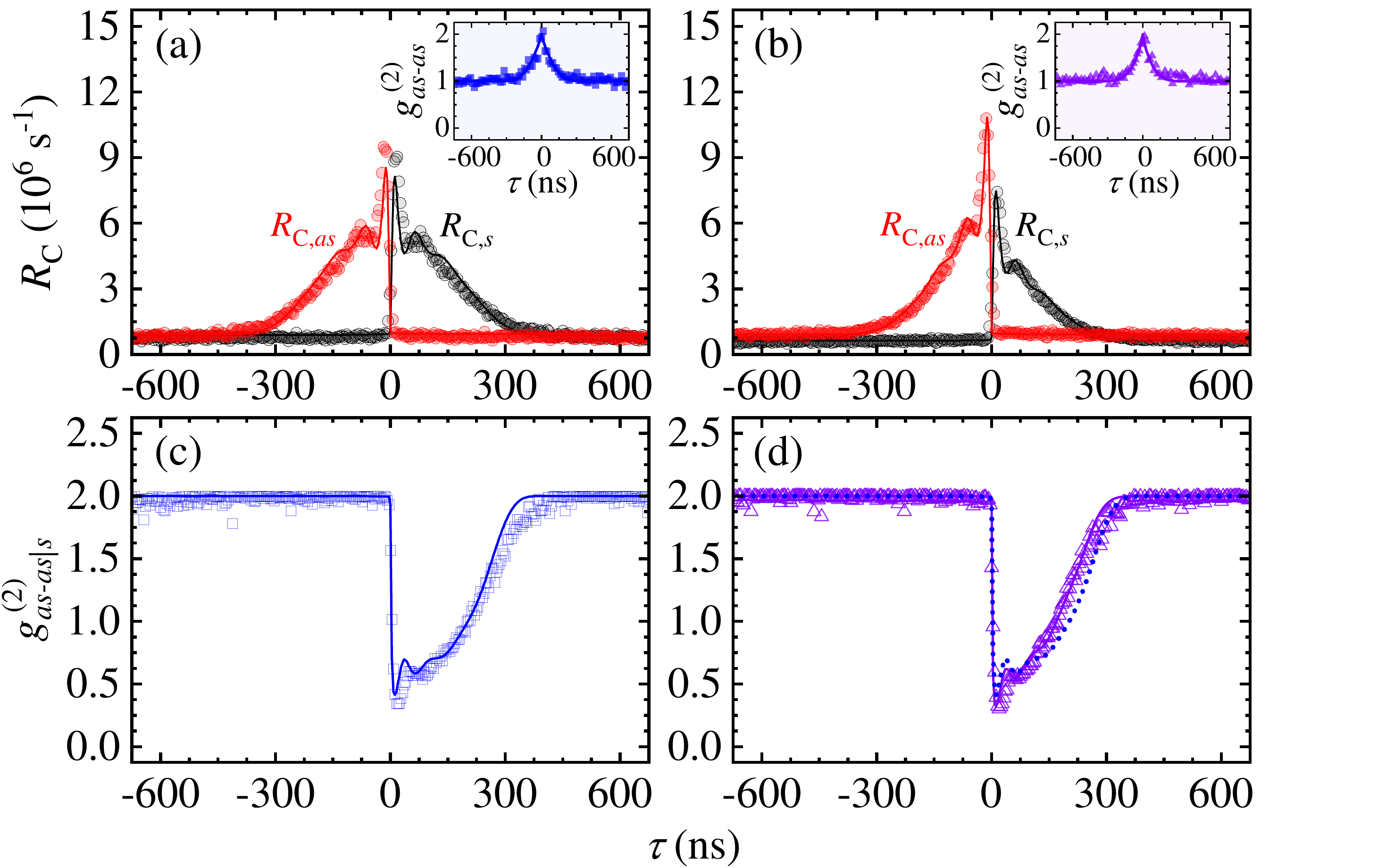}
  \caption
   {
     Asymmetric biphoton generation with higher optical depth (${\rm OD}=35$). Experimental parameters: $\Omega_d=\Omega_c=2\Gamma$, $\Delta_d=20\Gamma$, and $\Delta kL=0.37\pi$. All main data use a 4-ns time bin, while insets in (a) and (b) use a 16-ns time bin. (a) Coincidence count rates $R_{\rm C}$ with $\gamma_{21}=0.001\Gamma$. Experimental results for $R_{{\rm C},s}^{\rm exp}$ and $R_{{\rm C},as}^{\rm exp}$ (black and red circles, respectively) are derived from $N_{{\rm C},s}$ and $N_{{\rm C},as}$. The corresponding theoretical predictions $R_{{\rm C},s}^{\rm theo}$ and $R_{{\rm C},as}^{\rm theo}$ (black and red curves) are obtained from the theoretical model. In this case, the photon generation rates are $R_{s}=8.8 \times 10^5 \,\, \mathrm{s}^{-1}$ and $R_{as}=8.4 \times 10^5 \,\, \mathrm{s}^{-1}$. The inset shows the unconditional autocorrelation function $g_{as\text{-}as}^{(2)}$, where blue squares and the curve indicate the experimental data and theoretical prediction, respectively, measured with a 16-ns time bin. (b) $R_{\rm C}$ with $\gamma_{21}=0.1\Gamma$. The inset displays $g_{as\text{-}as}^{(2)}$ using purple triangles (experimental data) and the theoretical curve. Here, the photon generation rates are $R_{s}=8.7 \times 10^5 \,\, \mathrm{s}^{-1}$ and $R_{as}=6.0 \times 10^5 \,\, \mathrm{s}^{-1}$. (c) Conditional autocorrelation function $g_{as\text{-}as|s}^{(2)}$ for $\gamma_{21} = 0.001\Gamma$. The blue curve and squares represent the theoretical prediction and experimental data, respectively. (d) $g_{as\text{-}as|s}^{(2)}$ for $\gamma_{21} = 0.1\Gamma$. The purple curve and triangles represent the theoretical prediction and experimental data, respectively. The theoretical prediction for (c) is provided with the blue dotted curve in (d).
   }
  \label{fig4}
  \end{figure}
}


\section{Introduction}

Photons are essential quantum information carriers in networks due to their low decoherence, ease of manipulation, and long-distance propagation with minimal loss. These attributes make photons ideal for quantum communication and integral to quantum technologies, including quantum key distribution~\cite{QKD1, QKD2, QKD3, QKD4}, quantum computing~\cite{comp1, comp2, comp3}, and quantum teleportation~\cite{tele1, tele2, tele3}. Consequently, the development of reliable quantum light sources has become a critical area of research. Among these sources, biphotons are distinguished for their unique properties. They serve as heralded single-photon sources—where detecting one photon guarantees the presence of its counterpart—and offer versatility in generating diverse forms of entanglement. Beyond their intrinsic frequency-time entanglement~\cite{FT1, FT2, FT3, FT4}, biphotons enable tailored configurations for polarization-encoded~\cite{pol1, pol2, pol3, pol4} and orbital angular momentum (OAM)-encoded entanglement~\cite{OAM1, OAM2, OAM3}, providing multiple degrees of freedom for quantum information encoding.

Among the most widely used biphoton sources, spontaneous parametric down-conversion (SPDC)~\cite{SPDC1, SPDC2, SPDC3, SPDC4, SPDC5} and spontaneous four-wave mixing (SFWM)~\cite{SFWM1, SFWM2, SFWM3, SFWM4, SFWM5, SFWM6} each excel in different regimes. SPDC is well-suited for broadband applications, enabling faster computational and operational processes. In contrast, SFWM—a narrowband biphoton source—offers an extended coherence time and excellent compatibility with atomic systems, making it ideal for narrowband quantum applications. SFWM achieves a biphoton coherence time, $\tau_{\rm delay}$, exceeding 10 $\mu$s, which corresponds to a Fourier-limited bandwidth of 50 kHz~\cite{SFWM7}. This narrow bandwidth not only ensures long-distance phase coherence but also enhances compatibility with a wide variety of atomic systems~\cite{app1, app2, app3, app4, app5, app6, app7, app8}. Recent studies have further demonstrated its potential in applications such as dispersion cancellation within quantum interferometers~\cite{app9} and the exploration of which-path complementarity through induced and stimulated coherences~\cite{app10}. These advancements underscore the pivotal role of SFWM-based biphoton sources in driving progress in quantum information science.

To harness biphoton sources effectively, a deep understanding of their intrinsic properties—coherence, entanglement, and spectral characteristics—is crucial for advancing diverse applications. For SFWM, operating conditions closer to atomic resonance lead to intriguing phenomena such as superradiance~\cite{SR1, SR2, SR3, SR4}, oscillatory temporal correlations~\cite{osci1, osci2, osci3}, and optical precursors~\cite{pre1, pre2, pre3}. Furthermore, our previous work has experimentally demonstrated that photons generated via SFWM do not always form pairs~\cite{Shiu1, Shiu2}. This unpaired characteristic reduces with increasing optical depth (OD), where higher OD promotes photon pairing. At low OD, while higher pump power enhances photon generation rates, the fraction of paired photons remains significantly lower compared to high-OD conditions. Investigating this pairing ratio offers valuable insights into the fundamental properties of biphoton generation.

\figone 

In this study, we investigate asymmetric biphoton generation in double-$\Lambda$ SFWM~\cite{SFWM8, SFWM9, SFWM10, Kolchin, SFWM11, SFWM12, SFWM13, SFWM14, SFWM15} within a cold atomic ensemble. Specifically, we examine how ground-state decoherence and phase mismatch affect biphoton generation dynamics, leading to notable asymmetry in photon generation rates between the two output modes. This asymmetry is directly reflected in pairing ratios and coincidence count rates, revealing essential insights into generation processes. Moreover, we provide experimental results on both unconditional and conditional autocorrelation functions. Interestingly, while ground-state decoherence reduces the generation of temporally correlated photons, it paradoxically enhances the purity of heralded single photons. Understanding these intrinsic biphoton properties offers insights into generation dynamics and expands their applications in quantum technologies.


\section{Experimental Setup}

Figure~\ref{fig1}(a) illustrates the energy levels relevant to biphoton generation via double-$\Lambda$ SFWM in a cold $^{87}$Rb ensemble. Region I encompasses the laser cooling process using a magneto-optical trap (MOT), which involves trapping and repumping lasers. The trapping laser, addressing the transition $|5S_{1/2},\,F=2\rangle\leftrightarrow|5P_{3/2},\,F'=3\rangle$, is red-detuned by 27 MHz to achieve Doppler cooling, effectively cooling atoms in the $|5S_{1/2},\,F=2\rangle$ ground state. The repumping laser, addressing the transition $|5S_{1/2},\,F=1\rangle\leftrightarrow|5P_{3/2},\,F'=2\rangle$, transfers the atomic population from $F=1$ to $F=2$, allowing the trapping laser to interact with these atoms. The trapping laser is stabilized using an external cavity diode laser (ECDL). A portion of the ECDL output passes through a fiber-based electro-optic modulator (EOM), modulated at 6.8 GHz, to lock the repumping laser. These processes ensure the $^{87}$Rb atoms are cooled to approximately $300\,\mathrm{\mu K}$. Timing sequences, illustrated in Fig.~\ref{fig1}(b), are controlled using acousto-optical modulators. To minimize ground-state decoherence caused by residual magnetic fields, the MOT's magnetic field ($\textbf{B}$-field) is turned off 1500 $\mathrm{\mu s}$ before initiating the biphoton experiment (Region II). In addition, prior to biphoton generation, a $\sigma^+$-polarized coupling field, addressing the $|5S_{1/2},\,F=2\rangle \leftrightarrow |5P_{3/2},\,F'=2\rangle$ transition with Rabi frequency $\Omega_c$, prepares the atomic population in the ground state $F=1$. The resulting Zeeman state distribution is approximately 20\% in $m_{F=1}=-1$, and 40\% in both $m_{F=1}=0$ and $m_{F=1}=+1$.

In Region II, two optical fields—the driving field and the coupling field—simultaneously interact with cold atoms to induce SFWM, resulting in the generation of temporally correlated biphotons. The $\sigma^+$-polarized driving field, addressing the $|5S_{1/2},\,F=1\rangle\leftrightarrow|5P_{3/2},\,F'=2\rangle$ transition with a large detuning $\Delta_d$ and Rabi frequency $\Omega_d$, initiates spontaneous Raman scattering, during which Stokes photons are generated concurrently with the establishment of ground-state coherence. This coherence is subsequently converted into anti-Stokes photons by the coupling field, thereby completing the four-wave mixing (FWM) process. Synchronization of the driving and coupling fields is achieved through the ECDL. The coupling field is directly injected by the ECDL, while the driving field is phase-locked to the ECDL via a fiber-based EOM. Since both $\Omega_d$ and $\Omega_c$ are $\sigma^+$-polarized, population may accumulate in the dark state $|5S_{1/2},\,F=2,\,m_F=+2\rangle$, which does not interact with either field. To mitigate this, a 20-$\mathrm{\mu s}$ trapping sequence is applied after each 10-$\mathrm{\mu s}$ biphoton experiment to redistribute the ground-state population in $F=1$. This timing sequence operates at a repetition rate of 200 Hz, with each 5-ms cycle containing six 10-$\mathrm{\mu s}$ driving pulses, enhancing biphoton generation efficiency.

Figure~\ref{fig1}(c) illustrates the experimental setup. The backward configuration is employed, where $\Omega_d$ and $\Omega_c$ are counterpropagating. Modifications to the setup result in slight differences in parameters compared to previous experiments~\cite{Shiu1, Shiu2}. The $1/e^2$ beam widths of $\Omega_d$ and $\Omega_c$ are measured to be 250 and 730 $\mu$m, respectively, with powers of $7.5 \,\,\mathrm{\mu W}$ ($\Omega_d = 1\Gamma$) and $29.2 \,\,\mathrm{\mu W}$ ($\Omega_c = 1\Gamma$). Here, $\Gamma$ represents the spontaneous decay rate of the excited state $|3\rangle$ in Fig.~\ref{fig1}(a). Stokes and anti-Stokes photons are collected at angles of 1.7$^\circ$ relative to the driving and coupling fields, respectively. To suppress optical leakage, an etalon filter set (EFS) is employed, consisting of two etalons separated by an optical isolator. Each etalon provides an extinction ratio of approximately 30 dB and a bandwidth of around 100 MHz. For anti-Stokes photons, the detection channel is further split using a beam splitter in a Hanbury-Brown-Twiss (HBT) setup to enable autocorrelation measurements. The extinction ratios for the Stokes channel and the two anti-Stokes channels are measured to be 114 dB, 121 dB, and 124 dB, respectively.

Biphotons are detected using fiber-coupled single-photon counting modules (SPCM, AQRH-13-FC). Upon detection, the SPCM generates an 8-ns electrical signal, which is split via a BNC splitter and simultaneously sent to a digital oscilloscope (DO, RTO2014) and a time-of-flight multiscaler (TOF, MCS6A-4T8). The DO continuously monitors aggregate photon counts for each channel, enabling real-time tracking of the etalon's stability. Simultaneously, the TOF records precise time stamps of detected signals, organizing the data into three key components: detection cycle, signal arrival time within the cycle, and originating channel. This information is critical for analyzing coincidence counts across channels. When biphotons are generated as temporally correlated pairs, they are detected within the biphoton correlation time, forming a structured biphoton wavepacket that deviates from uniformity. This structured wavepacket provides crucial insights into the temporal and spectral characteristics of the biphoton source, serving as a basis for further optimization.


\section{Theoretical Framework}

\subsection{Atomic Response and Photon Generation Rates}

In the theoretical framework, we employ the Heisenberg--Langevin operator approach to characterize biphoton generation~\cite{Kolchin}. The propagating properties of the generated Stokes and anti-Stokes photons are described by the quantized field $\hat{E}_m=\sqrt{\frac{\hslash\bar{\omega}_m}{2\epsilon_0V}}e^{i(\vec{k}_m\cdot\vec{r}-\bar{\omega}_mt)}\int d\omega e^{-i\omega t}\widetilde{a}_m(z,\omega)+\sqrt{\frac{\hslash\bar{\omega}_m}{2\epsilon_0V}}e^{-i(\vec{k}_m\cdot\vec{r}-\bar{\omega}_mt)}\int d\omega e^{-i\omega t}\widetilde{a}_m^\dagger(z,-\omega)$, where the subscript $m = s$ and $m = as$ correspond to the Stokes and anti-Stokes fields, respectively. The pencil-shaped interaction volume $V$ is determined by the overlapping region of $\Omega_d$ and $\Omega_c$ within the atomic medium. The operators $\widetilde{a}_s(z,\omega)$ and $\widetilde{a}_{as}(z,\omega)$ obey the frequency-domain Maxwell–Schrödinger equations:
\begin{align}
&\left(
-\frac{i\omega}{c}+
\frac{\partial}{\partial z}
\right)\widetilde{a}_s
=\frac{iNg_s^*}{c}\widetilde{\sigma}_{23},\label{eq1}
\\
&
\left(
-\frac{i\omega}{c}-
\frac{\partial}{\partial z}+i\Delta k
\right)\widetilde{a}_{as}^\dagger
=-\frac{iNg_{as}}{c}\widetilde{\sigma}_{41}
.\label{eq2}
\end{align} 
Here, $N$ represents the total number of atoms, and $c$ denotes the speed of light. The interaction strength between the Stokes field and the atoms is characterized by the coupling constant $g_s = d_{32}\sqrt{\bar{\omega}_s/2\hslash\epsilon_0V}$, where $d_{32}$ denotes the dipole moment of the $|2\rangle\leftrightarrow|3\rangle$ transition and $\bar{\omega}_s$ is the central angular frequency of the Stokes field. The coupling constant between the anti-Stokes field and the atoms is similarly given by $g_{as} = d_{41}\sqrt{\bar{\omega}_{as} / 2\hslash\epsilon_0 V}$. The phase mismatch arising from the optical field configuration is characterized by the parameter $\Delta k = |\vec{k}_d - \vec{k}_s + \vec{k}_c - \vec{k}_{as}|$. Additionally, $\widetilde{\sigma}_{jk}$ is the frequency-domain collective atomic operator, defined as the Fourier transform of the time-domain operator $\hat{\sigma}_{jk}$, which satisfies the Heisenberg–Langevin equations: $\frac{\partial}{\partial t}\hat{\sigma}_{jk}=\frac{i}{\hslash}[\hat{H},\hat{\sigma}_{jk}]+\hat{R}_{jk}+\hat{F}_{jk}$. Here, the terms $\hat{R}_{jk}$ and $\hat{F}_{jk}$ denote the relaxation term and the Langevin noise operator, respectively. Therefore, $\widetilde{\sigma}_{jk}$ inherently includes both dispersion effects and ground-state decoherence. As a result of the above definitions and derivations, we obtain:
\begin{align}
&\begin{bmatrix}
	\widetilde{a}_s(L,\omega) \\ \widetilde{a}_{as}^\dagger(0,-\omega)
\end{bmatrix}
=
\begin{bmatrix}
	A(\omega) & B(\omega) \\ C(\omega) & D(\omega)
\end{bmatrix}
\begin{bmatrix}
	\widetilde{a}_s(0,\omega) \\ \widetilde{a}_{as}^\dagger(L,-\omega)
\end{bmatrix}
+\sqrt{\frac{N}{c}}
\sum_{jk}
\int_0^Ldz\,
\begin{bmatrix}
	P_{jk}(z,\omega) \\ Q_{jk}(z,\omega)
\end{bmatrix}\widetilde{F}_{jk}(z,\omega)
.\label{eq3}
\end{align}
The parameters $A$ and $D$ represent the mode-preserving coefficients, which encapsulate the self-susceptibilities of the Stokes and anti-Stokes fields, respectively. In contrast, $B$ and $C$ are mode-converting coefficients that capture the cross-coupling between the two fields and quantify how each influences the evolution of the other. Meanwhile, the parameters $P_{jk}$ and $Q_{jk}$ describe the injection of the frequency-domain Langevin noise operator $\widetilde{F}_{jk}$ into the system dynamics, coupling into the Stokes and anti-Stokes fields, respectively. These parameters encompass the key physical mechanisms of SFWM, including Raman scattering, the effect of  electromagnetically induced transparency (EIT)~\cite{EIT1,EIT2,EIT3}, as well as the influences of ground-state decoherence and phase mismatch. In other words, these effects are simultaneously manifested in both the Stokes and anti-Stokes fields. Further theoretical details are provided in Appendix A.

Using Eq. \eqref{eq3}, the commutation relation $[\widetilde{a}_m(z,\omega),\widetilde{a}_m^\dagger(z,-\omega')]=\frac{L}{2\pi c}\delta(\omega+\omega')$, and the delta-function correlation of white noise $\langle\widetilde{F}_{jk}(z,\omega)\widetilde{F}_{j'k'}(z',\omega')\rangle=\frac{L}{2\pi N}\mathcal{D}_{jk,j'k'}\delta(z-z')\delta(\omega-\omega')$, the photon generation rates $R_m = \frac{c}{L}\langle\hat{a}_m^\dagger\hat{a}_m\rangle$ for the Stokes and anti-Stokes fields can be expressed as:
\begin{align}
R_s
=&\int\frac{d\omega}{2\pi}
\left(
|B|^2
+\sum_{jk,j'k'}\int_0^Ldz\,
\hspace{0.1em}
P_{jk}^*\mathcal{D}_{jk^\dagger,j'k'}P_{j'k'}
\hspace{0.3em}
\right)
\equiv\int\frac{d\omega}{2\pi}\widetilde{R}_s(\omega),\label{eq4}
\\
R_{as}
=&\int\frac{d\omega}{2\pi}
\left(
|C|^2
+\sum_{jk,j'k'}\int_0^Ldz\,
Q_{jk}\mathcal{D}_{jk,j'k'^\dagger}Q_{j'k'}^*
\right)
\equiv\int\frac{d\omega}{2\pi}\widetilde{R}_{as}(\omega),\label{eq5}
\end{align}
where $\widetilde{R}_{s(as)}(\omega)$ represents the spectrum of the Stokes (anti-Stokes) field. The photon generation rate serves as a valuable indicator for probing the underlying physical mechanisms of SFWM. In the regime of dynamic equilibrium, Stokes and anti-Stokes photons are continuously generated, with their spontaneous emission occurring in random directions, reflecting the isotropic nature of the vacuum field. However, when a photon propagates along the elongated axis of the interaction volume, it may interact with nearby atoms and induce a stimulated FWM process. This directional mechanism leads to the generation of time-correlated photon pairs, which are characterized by the mode-converting coefficients $B$ and $C$. Although the mathematical forms of $B$ and $C$ differ (see Appendix A), their respective contributions to the spectral responses of the Stokes and anti-Stokes fields remain identical under both ideal and non-ideal conditions. This equivalence arises from the fact that $B$ and $C$ quantify the correlated photon-pair generation mediated by the stimulated FWM process, reflecting the inherent coupling between the Stokes and anti-Stokes photons in this interaction.

Nevertheless, directional propagation alone does not ensure subsequent interaction with the medium. Some photons, even if propagating along the axis of the interaction volume, may escape without inducing stimulated FWM and thus remain unpaired. These unpaired photons are associated with the diffusion coefficients $\mathcal{D}_{jk^\dagger,j'k'}$ in Eq.~\eqref{eq4} and $\mathcal{D}_{jk,j'k'^\dagger}$ in Eq.~\eqref{eq5}. Although these diffusion coefficients and the corresponding $P_{jk}$ and $Q_{ij}$ terms differ in their explicit forms for the two output modes, the contributions from $\sum\int_0^Ldz\,P_{jk}^*\mathcal{D}_{jk^\dagger,j'k'}P_{j'k'}$ and $\sum\int_0^Ldz\,Q_{jk}\mathcal{D}_{jk,j'k'^\dagger}Q_{j'k'}^*$ to the spectral responses are identical in the absence of phase mismatch and ground-state decoherence. The presence of phase mismatch and ground-state decoherence breaks this symmetry, as discussed in detail in Section~3.2.
	
These two distinct pathways—coherent pair production (i.e., photon-pair generation involving stimulated FWM) and incoherent emission (i.e., SFWM without pairing)—naturally give rise to the pairing ratios, defined as $r_{p,s} = R_s^{-1} \int \frac{d\omega}{2\pi} |B|^2$ and $r_{p,as} = R_{as}^{-1} \int \frac{d\omega}{2\pi} |C|^2$, which quantify the fraction of correlated photon generation. When both the phase mismatch and ground-state decoherence are small, these two pairing ratios become nearly identical. These ratios can be increased by enhancing the atomic ensemble density, thereby increasing the probability that the generated photons interact with nearby atoms and undergo stimulated FWM before leaving the medium~\cite{Shiu1}.

\subsection{Mechanisms of Asymmetric Biphoton Generation}

In an idealized setting that is free from both phase mismatch and ground-state decoherence, the photon generation rates $R_s$ and $R_{as}$ are identical. In practice, however, ground-state decoherence is inevitable, and phase mismatch is intrinsic to the backward-generation geometry. These imperfections reduce the efficiency with which spontaneously emitted photons can induce stimulated FWM in neighboring atoms. Although the mode-converting coefficients $B$ and $C$ are influenced by such non-idealities, their spectral contributions remain symmetric because both originate from the same stimulated FWM process. In contrast, the subsequent terms in Eqs.~\eqref{eq4} and \eqref{eq5} are more susceptible to asymmetries induced by phase mismatch and decoherence. This sensitivity arises from the fact that, under such conditions, resonant anti-Stokes photons experience greater losses than their off-resonant Stokes counterparts. Part of this imbalance is already incorporated into the mode-converting coefficients through their influence on the Stokes field. The remainder, which corresponds to photons that escape without undergoing stimulated interactions, directly contributes to the asymmetry in the overall generation rates. This asymmetry is manifested not only as a discrepancy between $R_s$ and $R_{as}$ but also as an imbalance between the pairing ratios $r_{p,s}$ and $r_{p,as}$. The implications of these asymmetries will be further examined in the following discussion of experimental results.

To assess the impact of asymmetric biphoton generation on temporal correlations, one may turn to the normalized cross-correlation function. This function is a key metric for characterizing non-classical temporal correlations between two fields. It is expressed as:
\begin{align}
g_{s\text{-}as}^{(2)}(\tau)
=\frac{\langle\hat{a}_{s}^\dagger(t)\hat{a}_{as}^\dagger(t+\tau)\hat{a}_{as}(t+\tau)\hat{a}_{s}(t)\rangle}{\langle\hat{a}_{s}^\dagger(t)\hat{a}_{s}(t)\rangle\langle\hat{a}_{as}^\dagger(t+\tau)\hat{a}_{as}(t+\tau)\rangle}
=
1+
\frac{1}{R_sR_{as}}
\left|
\int \frac{d\omega}{2\pi}
e^{-i\omega\tau}
\left(
\vphantom{\frac{A}{B}}
B^*D
+
\sum_{jk,j'k'}\int_0^Ldz
\vphantom{\frac{A}{B}}
P_{jk}^*\mathcal{D}_{jk^\dagger,j'k'}Q_{j'k'}\right)
\right|^2.\label{eq6}
\end{align}
The constant background level of 1 in Eq. \eqref{eq6} includes contributions from the uncorrelated nature of distinct biphoton pairs as well as from photons that do not participate in the stimulated FWM process. By contrast, the subsequent term represents the temporal quantum correlations of biphotons established in a specific propagation direction through the stimulated FWM process. These correlations correspond to the biphoton wavepacket, which describes the wavepacket of anti-Stokes single photons when Stokes photons are post-selected as triggers. In biphoton sources, the background level of 1 defines the maximum signal-to-background ratio, given by $r_{\rm SB} \equiv \max[g_{s\text{-}as}^{(2)}(\tau) - 1]$. A larger value of $r_{\rm SB}$ indicates a more pronounced deviation of biphoton behavior from classical predictions. The Cauchy–Schwarz inequality, given by $[g_{s\text{-}as}^{(2)}(\tau)]^2 \leq g_{s\text{-}s}^{(2)}(0)g_{as\text{-}as}^{(2)}(0)$, is a widely used criterion for identifying quantum correlations between two fields. A violation of this inequality signifies the presence of nonclassical correlations, and larger $r_{\rm SB}$ values correspond to a greater deviation from the classical bound of unity. Additionally, the temporal profile of the Stokes photon wavepacket, conditioned on the detection of an anti-Stokes photon, is commonly described by $g_{as\text{-}s}^{(2)}(\tau) = g_{s\text{-}as}^{(2)}(-\tau)$. Although both $g_{s\text{-}as}^{(2)}(\tau)$ and $g_{as\text{-}s}^{(2)}(\tau)$ are affected by phase mismatch and ground-state decoherence, their normalization to a background level of 1 suppresses observable asymmetries in the biphoton generation process. As a result, this conventional formulation is incapable of revealing the asymmetry in photon generation rates between the Stokes and anti-Stokes fields.

To expose the asymmetry in biphoton generation using the cross-correlation function, it is instructive to examine the coincidence count rate $R_{\rm C}(\tau) = R_s R_{as} g_{s\text{-}as}^{(2)}(\tau) \Delta T$, which represents the anti-Stokes photon count rate conditioned on detecting a specific number of Stokes photons within the time interval $\Delta T$. When the time interval is set to the inverse of the Stokes photon generation rate, i.e., $\Delta T = R_s^{-1}$, the number of Stokes photons becomes exactly 1, and the coincidence count rate simplifies to $R_{{\rm C},s}(\tau) = R_{as} g_{s\text{-}as}^{(2)}(\tau)$. In this case, the average number of anti-Stokes photons temporally correlated with a single Stokes photon is represented by the pairing ratio of a Stokes photon, $r_{p,s}$. This ratio corresponds to the area of the temporally correlated biphoton wavepacket, while the uncorrelated background rate equals the anti-Stokes photon generation rate $R_{as}$. Analogously, the coincidence count rate $R_{{\rm C},as}(\tau) = R_s g_{as\text{-}s}^{(2)}(\tau)$ characterizes the Stokes photon wavepacket conditioned on the detection of an anti-Stokes photon. The pairing ratio $r_{p,as}$ corresponds to the area under the biphoton wavepacket, while the uncorrelated background level is given by the Stokes photon generation rate $R_s$. This formulation makes explicit the roles of the photon generation rates $R_s$ and $R_{as}$, together with the pairing ratios $r_{p,s}$ and $r_{p,as}$, in revealing the asymmetry in biphoton generation caused by phase mismatch and ground-state decoherence.

\subsection{Temporal Correlations and Autocorrelation Functions}

In addition to analyzing the cross-correlation function, we also examine the influence of phase mismatch and ground-state decoherence on the normalized autocorrelation function~\cite{Scully}. By applying Eq.~\eqref{eq3} and the generalized Wick's theorem~\cite{Wick}, the autocorrelation function can be recast as $g_{m\text{-}m}^{(2)}(\tau)=1+R_m^{-2}|\int \frac{d\omega}{2\pi} e^{-i\omega\tau}\widetilde{R}_m|^2$. At zero delay, one obtains $g_{s\text{-}s}^{(2)}(0) = g_{as\text{-}as}^{(2)}(0) = 2$, which confirms that the generated Stokes and anti-Stokes photons each follow a thermal field distribution. Moreover, since both phase mismatch and ground-state decoherence affect $\widetilde{R}_m$, their influence manifests in the coherence times of the Stokes and anti-Stokes fields.

Although the autocorrelation function described above is a standard diagnostic tool in quantum optics, it is based solely on measurements of a single optical field. Consequently, it neglects correlations between the Stokes and anti-Stokes photons, and is therefore termed the unconditional autocorrelation function. In contrast, biphoton systems focus primarily on the photon statistics of one field conditioned on detecting its time-correlated partner, known as the conditional autocorrelation function. This measurement involves three temporal variables: the detection time of a Stokes photon and the two possible detection times of an anti-Stokes photon at the outputs of the HBT setup. To assess the heralded-state purity of anti-Stokes photons as a function of delay relative to their paired Stokes photons, we set the HBT delay time to zero. This setting probes the probability of detecting an anti-Stokes photon in coincidence with a preceding Stokes photon, thereby revealing the single-photon statistics of the anti-Stokes field conditioned on Stokes detection. Utilizing the generalized Wick's theorem, the zero-delay conditional autocorrelation function can be obtained as follows:
\begin{align}
g_{as\text{-}as|s}^{(2)}(\tau)
=
\frac{\langle\hat{a}_s^\dagger(t)\hat{a}_s(t)\rangle
\langle\hat{a}_s^\dagger(t)\hat{a}_{as,1}^\dagger(t+\tau)\hat{a}_{as,2}^\dagger(t+\tau)\hat{a}_{as,2}(t+\tau)\hat{a}_{as,1}(t+\tau)\hat{a}_s(t)\rangle}
{\langle\hat{a}_s^\dagger(t)\hat{a}_{as,1}^\dagger(t+\tau)\hat{a}_{as,1}(t+\tau)\hat{a}_s(t)\rangle\langle\hat{a}_s^\dagger(t)\hat{a}_{as,2}^\dagger(t+\tau)\hat{a}_{as,2}(t+\tau)\hat{a}_s(t)\rangle}
=
\frac
{4g_{s\text{-}as}^{(2)}(\tau)-2}
{\left[g_{s\text{-}as}^{(2)}(\tau)\right]^2}.\label{eq7}
\end{align}
Equation \eqref{eq7} shows that the zero-delay conditional autocorrelation function of heralded anti-Stokes photons is directly linked to the biphoton cross-correlation function. Further details of the theoretical derivation are presented in Appendix B. As the purity of the biphotons increases, the heralded anti-Stokes photons exhibit behavior increasingly similar to that of an ideal Fock state, with $g_{as\text{-}as|s}^{(2)}(\tau)$ approaching zero. This underscores the importance of $g_{s\text{-}as}^{(2)}(\tau)$, which is directly related to $r_{\rm SB}$, as a key parameter for characterizing and potentially optimizing biphoton sources. Moreover, as $\tau$ increases, the temporal correlation between Stokes and anti-Stokes photons gradually diminishes, causing $g_{s\text{-}as}^{(2)}(\tau)$ to approach unity. In this regime, $g_{as\text{-}as|s}^{(2)}(\tau)$ approaches 2, indicating that the detected Stokes and anti-Stokes photons originate from different photon pairs. Consequently, the Stokes photon detection no longer influences the thermal distribution of independently emitted anti-Stokes photons.


\figtwo 

\section{Results and Discussion}

\subsection{Analysis of Coincidence Count Rates}

Figure~\ref{fig2}(a) presents the raw experimental data obtained with the following parameters: an optical depth of ${\rm OD}=20$, $\Omega_d = \Omega_c = 2\Gamma$, $\Delta_d = 20\Gamma$, and a ground-state decoherence rate of $\gamma_{21} = 0.001\Gamma$. The phase mismatch of $\Delta kL = 0.37\pi$ is calculated based on the separation angle of $1.7^\circ$ and the medium length of $L = 4$ mm. The time bin size is set to $\Delta\tau$ = 4 ns, and the total data acquisition time is 1 hour. The black curve represents the coincidence counts using Stokes photons as triggers, while the red curve corresponds to anti-Stokes photons as triggers. These two curves exhibit near-perfect symmetry, differing only due to time-reversal symmetry, where one curve is the mirror image of the other with respect to the time axis~\cite{SFWM15}.

The symmetry observed in the coincidence count data arises from the fact that the TOF system records all signals from the SPCMs, including both correlated and uncorrelated events. Consequently, the choice of trigger channel affects only the sign of the recorded time difference, not its magnitude (i.e., for the $i$-th event, $\tau_{1,i} = t_{as,i} - t_{s,i}$, or equivalently, $\tau_{2,i} = t_{s,i} - t_{as,i} = -\tau_{1,i}$). While this representation captures certain features of the biphoton wavepacket, it does not fully reveal the underlying physics, particularly when the generation rates of Stokes and anti-Stokes photons differ, i.e., $R_s \neq R_{as}$. These discrepancies in generation rates are primarily attributed to ground-state decoherence and phase mismatch, with $\gamma_{21}$ and $\Delta kL$ being key factors. To more effectively illustrate the characteristics of biphotons, it is more suitable to present the experimental results in terms of the coincidence count rate~\cite{SFWM8, Shiu1, Shiu2, Kolchin}. This approach not only directly highlights the temporal correlations between photons but also accounts for differences in photon generation rates, providing a clearer understanding of biphoton dynamics.

Figure~\ref{fig2}(b) displays the coincidence count rates obtained by processing the raw experimental data. The results are classified according to the chosen trigger channel, denoted as $R_{{\rm C},s}$ for the Stokes trigger channel and $R_{{\rm C},as}$ for the anti-Stokes trigger channel. To facilitate a clearer discussion, we adopt superscript notation. The experimentally measured coincidence count rates are represented by circles, with $R_{{\rm C},s}^{\rm exp}$ shown in black and $R_{{\rm C},as}^{\rm exp}$ in red. The corresponding theoretical predictions are depicted as curves, labeled as $R_{{\rm C},s}^{\rm theo}$ and $R_{{\rm C},as}^{\rm theo}$. These theoretical rates include contributions from environmental noise, such as optical leakage and dark counts from the SPCMs. These noise sources primarily contribute to the uncorrelated background, which is added to the noise-free theoretical rates, $R_{{\rm C},s}$ and $R_{{\rm C},as}$. The total theoretical rates are expressed as $R_{{\rm C},s}^{\rm theo} = R_{{\rm C},s} + R_{{\rm env},s}$ and $R_{{\rm C},as}^{\rm theo} = R_{{\rm C},as} + R_{{\rm env},as}$, where $R_{{\rm env},s}$ and $R_{{\rm env},as}$ represent the contributions from environmental noise. To further analyze the coincidence count rate, we consider the case where Stokes photons are used as triggers. The relationship between $R_{{\rm C},s}$ and the coincidence counts $N_{{\rm C},s}$ is expressed as:
\begin{align}
N_{{\rm C},s}=
N_sP_sR_{{\rm C},s}\eta_{as}\Delta\tau
+N_s(1-P_s)(R_{as}\eta_{as}+R_{\rm noise}^{as})\Delta\tau
+N_sP_sR_{\rm noise}^{as}\Delta\tau
,\label{eq8}
\end{align}
where $N_s$ represents the total received counts in the Stokes channel, comprising both Stokes photons and environmental noise. In our prior works~\cite{Shiu1, Shiu2}, $N_s$ was consistently set to $2^{18}$. However, as the objective of this study differs, we adopted a different approach by fixing the duration of biphoton collection. The noise count rates, $R_{\rm noise}^{s}$ in the Stokes channel and $R_{\rm noise}^{as}$ in the anti-Stokes channel, were determined by turning off the $\textbf{B}$-field of the MOT. These rates, approximately 600 and 570 counts/s, respectively, are attributed to environmental noise. When the $\textbf{B}$-field is turned on, the received count rate in the Stokes channel becomes $R_s \eta_s + R_{\rm noise}^{s}$. The purity $P_s$ in the Stokes channel is determined by $\frac{R_s\eta_s}{R_s\eta_s + R_{\rm noise}^{s}}$, where $\eta_s$ denotes the collection efficiency of Stokes photons. Photon reception in the anti-Stokes channel is described analogously. For the experimental data presented in Fig.~\ref{fig2}(b), the total received count in the Stokes channel is $N_s = 607609$, and the purity $P_s$ is approximately 95.4\%, indicating that $N_s P_s = 579659$ Stokes photons were received over a collection time of 1 hour. Therefore, the first term in Eq. \eqref{eq8} represents the coincidence counts arising from the reception of both Stokes and anti-Stokes photons. This analysis implies that, within the total Stokes photon counts $N_s P_s$, each individual Stokes photon count corresponds to an anti-Stokes photon count rate of $R_{{\rm C},s} \eta_{as}$. The second and third terms in the coincidence count rate describe background contributions. The second term corresponds to $N_s(1-P_s)$ environmental noise counts in the Stokes channel, which contribute only to the uncorrelated background regardless of anti-Stokes channel detections. The third term, in contrast, reflects the fact that, even though $N_s P_s$ Stokes photons are received, they cannot exhibit temporal correlation with $R_{\rm noise}^{as}$.

The collection efficiencies $\eta_s = 2.6\%$ and $\eta_{as} = 2.1\%$ in Fig.~\ref{fig2}(b) are determined based on the theoretical generation rates $R_s = 4.8 \times 10^5 \,\, \mathrm{s}^{-1}$ and $R_{as} = 4.5 \times 10^5 \,\, \mathrm{s}^{-1}$. These values are influenced by both the temperature stability of the EFS and variations in optical alignment, resulting in slight day-to-day fluctuations. Nevertheless, $\eta_s$ and $\eta_{as}$ consistently remain within the range of $2\%$ to $3\%$. Evidence supporting this method includes the observed agreement between the experimental biphoton wavepacket and the signal-to-background ratio $r_{\rm SB}$ with theoretical predictions, demonstrating the validity of our approach. The background in Eq. \eqref{eq8}, $N_s(R_{as}\eta_{as} + R_{\rm noise}^{as})\Delta\tau$, is directly measurable from the experiment, independent of theoretical predictions. Incorrect parameter estimation would cause deviations in the biphoton wavepacket shape and discrepancies in $r_{\rm SB}$ compared with theoretical predictions. Notably, such discrepancies are absent in our data. Consequently, the collection efficiencies determined by this method align with the experimental observations, demonstrating the reliability of our approach. To calculate the coincidence count rate, we divide Eq. \eqref{eq8} by $N_sP_s\eta_{as}\Delta\tau$. This yields the coincidence count rate, including environmental noise, expressed as $R_{{\rm C},s}^{\rm theo} = R_{{\rm C},s} + R_{{\rm env},s}$. Here, $R_{{\rm env},s} = \frac{1-P_s}{P_s}R_{as}+\frac{R_{\rm noise}^{as}}{P_s\eta_{as}}$, which represents the environmental noise contribution originating from both channels. This contribution is distinct from $R_{\rm noise}^{s}$ and $R_{\rm noise}^{as}$, which denote the independently measured noise count rates within the Stokes and anti-Stokes channels, respectively.

\renewcommand{\arraystretch}{1.5} 
\setlength{\tabcolsep}{10pt} 
\begin{table*}[t]
\centering
\begin{tabular}{c c | c c c c c c}
\hline\hline
OD & $\gamma_{21}$ $\left[\Gamma\right]$ & 
$R_s$ $\left[\mathrm{s}^{-1}\right]$ & $R_sr_{p,s}$ $\left[\mathrm{s}^{-1}\right]$ & $r_{p,s}$ & $R_{as}$ $\left[\mathrm{s}^{-1}\right]$ & $R_{as}r_{p,as}$ $\left[\mathrm{s}^{-1}\right]$ & $r_{p,as}$ \\
\hline 
20 & 0.001 & 
$4.8\times10^5$ & $3.3\times10^5$ & 0.70 & 
$4.5\times10^5$ & $3.3\times10^5$ & 0.74
\\
\hline
20 & 0.1 & 
$4.8\times10^5$ & $2.7\times10^5$ & 0.56 & 
$3.8\times10^5$ & $2.7\times10^5$ & 0.71
\\
\hline
35 & 0.001 & 
$8.8\times10^5$ & $6.7\times10^5$ & 0.77 & 
$8.4\times10^5$ & $6.7\times10^5$ & 0.81
\\
\hline
35 & 0.1 & 
$8.7\times10^5$ & $4.6\times10^5$ & 0.54 & 
$6.0\times10^5$ & $4.6\times10^5$ & 0.78
\\
\hline\hline
\end{tabular}
\caption{Theoretical predictions for the photon generation rate and pairing ratio based on the experimental conditions shown in Fig. \ref{fig2}, Fig. \ref{fig3}, and Fig. \ref{fig4} of this study. These predictions are calculated using Eqs. \eqref{eq6} and \eqref{eq7}. Under these conditions, the fixed experimental parameters are $\Omega_d = \Omega_c = 2\Gamma$, $\Delta_d = 20\Gamma$, and $\Delta kL = 0.37\pi$.}
\label{table1}
\end{table*}

\subsection{Effects of Phase Mismatch on Biphoton Dynamics}

In Fig.~\ref{fig2}(b), the experimental anti-Stokes photon generation rate, $R_{as}^{\rm exp} = 4.3 \times 10^5 \,\, \mathrm{s}^{-1}$, is determined by subtracting the environmental rate, $R_{{\rm env},s} = 0.5 \times 10^5 \,\, \mathrm{s}^{-1}$, from the background of $R_{{\rm C},s}^{\rm exp}$, which is $4.8 \times 10^5 \,\, \mathrm{s}^{-1}$. The experimental pairing ratio $r_{p,s}^{\rm exp} = 0.73$ is given by integrating the area under the correlated biphoton wavepacket, i.e., $r_{p,s}^{\rm exp} = \int_0^\infty [R_{{\rm C},s}^{\rm exp}(\tau) - R_{as}^{\rm exp} - R_{{\rm env},s}] d\tau$~\cite{Shiu1}. A non-unity pairing ratio indicates that post-selecting a Stokes photon does not always guarantee the simultaneous generation of an anti-Stokes photon. This phenomenon arises because not all photons generated within the atomic ensemble interact with neighboring atoms to undergo stimulated FWM. Photons that do not participate in stimulated FWM cannot establish temporally correlated biphotons in a specific direction, thereby preventing $r_p$ from reaching unity. The theoretical predictions for this study are summarized in Table~\ref{table1}, showing good agreement with the experimental results.

When anti-Stokes photons are selected as triggers, the experimental results in Fig.~\ref{fig2}(b) yield $R_s^{\rm exp} = 4.6 \times 10^5 \,\, \mathrm{s}^{-1}$ and $r_{p,as}^{\rm exp} = 0.77$. By combining the information from $R_{{\rm C},s}^{\rm exp}$ and $R_{{\rm C},as}^{\rm exp}$, the photon rate associated with stimulated FWM is determined to be $R_s^{\rm exp} r_{p,s}^{\rm exp} = R_{as}^{\rm exp} r_{p,as}^{\rm exp} = 3.3 \times 10^5 \,\, \mathrm{s}^{-1}$. This equality arises because Stokes and anti-Stokes photons mutually influence each other through stimulated FWM, resulting in a dynamic equilibrium. Photons not involved in stimulated FWM represent those generated without interacting with neighboring atoms. For Stokes photons, this rate is $R_s^{\rm exp}(1 - r_{p,s}^{\rm exp}) = 1.3 \times 10^5 \,\, \mathrm{s}^{-1}$, while for anti-Stokes photons, it is $R_{as}^{\rm exp}(1 - r_{p,as}^{\rm exp}) = 1.0 \times 10^5 \,\, \mathrm{s}^{-1}$.

With a small decoherence rate of $\gamma_{21} = 0.001\Gamma$ in Fig.~\ref{fig2}(b), the slight asymmetry between $R_s^{\rm exp}$ and $R_{as}^{\rm exp}$ is primarily attributed to the phase mismatch introduced by the backward configuration of the experimental setup. In an ideal phase-matched scenario ($\Delta kL = 0$), the theoretical photon rates involved in stimulated FWM are $R_s r_{p,s} = R_{as} r_{p,as} = 3.6 \times 10^5 \,\, \mathrm{s}^{-1}$, while the rates for photons not participating in stimulated FWM, $R_s(1 - r_{p,s})$ and $R_{as}(1 - r_{p,as})$, are $1.2 \times 10^5 \,\, \mathrm{s}^{-1}$. Compared to the asymmetry observed at $\Delta kL = 0.37\pi$, where the biphoton generation asymmetry is approximately $7\%$, the asymmetry in the phase-matched case ($\Delta kL = 0$) is minimal, with only about a $1\%$ residual difference. This remaining asymmetry is attributed to the non-zero ground-state decoherence.

The influence of phase mismatch on SFWM primarily arises from its interference with the interaction between the Stokes and anti-Stokes photons during the stimulated FWM process. The Stokes and anti-Stokes photons are generated via spontaneous Raman scattering, driven by the interaction of a far-detuned driving field and a resonant coupling field with the atomic ensemble. This generation process itself does not rely on phase matching. However, phase mismatch weakens the stimulated FWM effect between the Stokes and anti-Stokes photons and neighboring atoms, resulting in a reduction in the number of temporally correlated photon pairs, as reflected by the decreased pairing ratios $r_{p,s}$ and $r_{p,as}$. This weakened stimulated FWM effect further reduces the number of Stokes and anti-Stokes photons due to absorption by neighboring atoms, especially for $R_{as}$, which is influenced by the resonant anti-Stokes photons. In contrast, the generation of Stokes photons is far detuned from resonance, so $R_s$ remains nearly unaffected by phase mismatch under low OD conditions.

\subsection{Measurement and Analysis of Autocorrelation Functions}

Alongside the cross-correlation function, we measure the unconditional autocorrelation function of the anti-Stokes field, $g_{as\text{-}as}^{(2)}(\tau)$, as shown in Fig.~\ref{fig2}(c). The time bin is set to $\Delta\tau = 16 \,\, \mathrm{ns}$. The blue squares represent the experimental data, while the blue curve corresponds to theoretical calculations. The experimental data are accumulated over 8 hours. Due to the sensitivity of the cross-correlation function's optical precursor and oscillatory structure to experimental parameters, we do not present the cross-correlation function based on the results accumulated over 8 hours. At zero delay, $g_{as\text{-}as}^{(2)}(\tau = 0) \approx 2$, confirming that the generated photons follow a thermal state distribution. This distribution is characterized by a $1/e$ coherence time of $\tau_c = 115 \,\, \mathrm{ns}$, which reflects the intrinsic coherence time of the anti-Stokes field and is independent of measurements in the Stokes channel. This coherence time, significantly longer than the spontaneous emission time $\tau_{\rm spon} = \Gamma^{-1} \approx 26.5 \,\, \mathrm{ns}$, serves as evidence of the stimulated process occurring. Furthermore, it is controllable by manipulating the conditions of the intrinsic $\Lambda$-type EIT. As the time increases, the intensity correlation of anti-Stokes photons gradually decreases. The background value of 1 in the HBT measurement indicates that photons detected by the two channels originate from different photon wavepackets and, as a result, do not exhibit intensity correlation.

Beyond $g_{as\text{-}as}^{(2)}(\tau)$, we also demonstrate the conditional autocorrelation function in Fig.~\ref{fig2}(d). This function is derived using the conversion $g_{as\text{-}as|s}^{(2)}(\tau) = [4g_{s\text{-}as}^{(2)}(\tau) - 2][g_{s\text{-}as}^{(2)}(\tau)]^{-2}$. This result is determined solely from cross-correlation measurements, as it does not require directly measuring the time difference between the two anti-Stokes channels. Although we obtain $g_{as\text{-}as|s}^{(2)}(\tau)$ through this conversion without direct experimental verification (due to the extended measurement time exceeding system stability), a recent study in a hot atomic system experimentally confirms this relationship~\cite{SFWM16}. Furthermore, the conversion derived using the Heisenberg--Langevin operator approach aligns well with the results obtained in this hot atomic system. Notably, our theoretical framework, which describes a complete open quantum system, offers simpler and more direct expressions.

In Fig.~\ref{fig2}(b), the measured peak cross-correlation function at $\tau = 20 \,\, \mathrm{ns}$ is approximately 23. The corresponding  minimum value of $g_{as\text{-}as|s}^{(2)}(\tau)$ is estimated to be around 0.17, as shown in Fig.~\ref{fig2}(d). This value can be further reduced by decreasing the photon generation rate and biphoton correlation time. In our previous work~\cite{Shiu1}, we reported high-purity biphotons with $r_{\rm SB} = 241$ at a photon generation rate of $R_{as} = 5.0 \times 10^4 \,\, \mathrm{s}^{-1}$. Under these conditions, the minimum value of $g_{as\text{-}as|s}^{(2)}(\tau)$ was estimated to be approximately 0.016. Furthermore, with fixed photon generation rates and biphoton wavepacket bandwidth, improving the pairing ratio provides an alternative approach to enhance $r_{\rm SB}$.

\figthree

\subsection{Influence of Ground-State Decoherence on Biphoton Generation}

To further emphasize the characteristics of asymmetric biphoton generation, we examine an extreme condition to amplify the difference between $R_s$ and $R_{as}$. Because the phase mismatch in our experimental setup is not easily adjustable, we vary the ground-state decoherence rate instead. By applying an external transverse magnetic field during biphoton generation, we increase the ground-state decoherence rate to $\gamma_{21} = 0.1\,\Gamma$. The resulting coincidence counts are shown in Fig.~\ref{fig3}(a). Due to the reduction in temporally correlated photons, the observed coincidence counts are significantly lower than those in Fig.~\ref{fig2}(a), where $\gamma_{21} = 0.001\,\Gamma$. The corresponding coincidence count rates, $R_{{\rm C},s}^{\rm exp}$ and $R_{{\rm C},as}^{\rm exp}$, are presented in Fig.~\ref{fig3}(b). From these rates, we determine that $R_s^{\rm exp} = 4.8 \times 10^5 \,\,\mathrm{s}^{-1}$, $R_{as}^{\rm exp} = 3.8 \times 10^5 \,\,\mathrm{s}^{-1}$, $r_{p,s}^{\rm exp} = 0.59$, and $r_{p,as}^{\rm exp} = 0.75$.

We calculate the photon rate associated with stimulated FWM as $R_s^{\rm exp} r_{p,s}^{\rm exp} = R_{as}^{\rm exp} r_{p,as}^{\rm exp} = 2.8 \times 10^5 \,\,\mathrm{s}^{-1}$. In contrast, photons not participating in stimulated FWM are characterized by $R_s^{\rm exp}(1 - r_{p,s}^{\rm exp}) = 2.0 \times 10^5 \,\,\mathrm{s}^{-1}$ and $R_{as}^{\rm exp}(1 - r_{p,as}^{\rm exp}) = 1.0 \times 10^5 \,\,\mathrm{s}^{-1}$. A recent theoretical work with a similar energy level structure also predicted this phenomenon~\cite{SFWM17}. In that study, despite the presence of an additional external optical field coupling $|3\rangle$ to a Rydberg state $|5\rangle$, the theory predicts that an increase in $\gamma_{21}$ leads to asymmetric photon generation in the two output channels. This asymmetry arises from the contribution of photons not involved in stimulated FWM. Interestingly, although $R_{as}^{\rm exp}$ is substantially reduced compared to the $\gamma_{21} = 0.001\Gamma$ case, $r_{p,as}^{\rm exp}$ exhibits only a slight change. This occurs because $\gamma_{21}$ uniformly affects the loss of resonant anti-Stokes photons, regardless of their participation in the stimulated FWM. Conversely, due to the far off-resonant nature of Stokes photons, $R_s^{\rm exp}$ is minimally influenced by $\gamma_{21}$, thus remaining nearly constant. However, $\gamma_{21}$ still disrupts the stimulated FWM process, reducing the ability of the generated Stokes photons to engage in FWM with neighboring atoms and causing a notable drop in $r_{p,s}^{\rm exp}$. Consequently, the number of photons not involved in stimulated FWM increases significantly. In summary, $\gamma_{21}$ reduces $R_{as}$ and $r_{p,s}$ while its impact on $R_s$ and $r_{p,as}$ is minimal. This mechanism is fundamentally similar to the asymmetry caused by phase mismatch mentioned earlier.

In addition to influencing photon generation rates and pairing ratios, $\gamma_{21}$ also affects the biphoton coherence time $\tau_{\rm delay}$ and the anti-Stokes coherence time $\tau_c$, as reflected in $g_{s\text{-}as}^{(2)}(\tau)$ and $g_{as\text{-}as}^{(2)}(\tau)$, respectively. To quantify the impact of $\gamma_{21}$ on $\tau_{\rm delay}$, a suitable characteristic time must first be defined. In hot-atom systems, the most commonly used metric is the $1/e$ decay time, which is effective due to Doppler broadening that renders the biphoton wavepacket approximately exponential. However, in cold-atom systems, where atomic motion is negligible, atomic dispersion leads to more diverse biphoton wavepacket profiles, making the $1/e$ decay time a less reliable measure of coherence. Instead, we adopt a cumulative distribution function approach to define $\tau_{\rm delay}$~\cite{Shiu2}. From Fig.~\ref{fig2}(b), we determine that $\tau_{\rm delay} \approx 81 \,\, \mathrm{ns}$ at $\gamma_{21} = 0.001\Gamma$, while Fig.~\ref{fig3}(b) shows that $\tau_{\rm delay} \approx 70 \,\, \mathrm{ns}$ at $\gamma_{21} = 0.1\Gamma$. A similar analysis reveals that in Fig.~\ref{fig2}(c), $\tau_c \approx 115 \,\, \mathrm{ns}$ at $\gamma_{21} = 0.001\Gamma$, whereas Fig.~\ref{fig3}(c) indicates that $\tau_c \approx 96 \,\, \mathrm{ns}$ at $\gamma_{21} = 0.1\Gamma$. These results indicate that both $\tau_{\rm delay}$ and $\tau_c$ decrease as $\gamma_{21}$ increases, leading to the observed differences in $r_{\rm SB}$ ($r_{\rm SB} \approx 22$ for $\gamma_{21} = 0.001\Gamma$ and $r_{\rm SB} \approx 25$ for $\gamma_{21} = 0.1\Gamma$). This behavior can be attributed to changes in the coincidence count rate of $R_{{\rm C},as}$. While $\gamma_{21}$ reduces $\tau_{\rm delay}$, it has minimal impact on $r_{p,as}$ (the correlated area) and $R_s$ (the uncorrelated background). This effectively stretches the wavepacket vertically, enhancing $r_{\rm SB}$. Such an outcome is both intriguing and counterintuitive. These effects are also evident in $g_{as\text{-}as|s}^{(2)}(\tau)$, as shown in Figs.~\ref{fig2}(d) and \ref{fig3}(d). The minimum value of  $g_{as\text{-}as|s}^{(2)}(\tau)$ is approximately 0.17 and 0.15 for $\gamma_{21} = 0.001\Gamma$  and $\gamma_{21} = 0.1\Gamma$, respectively, demonstrating that an increase in $\gamma_{21}$ lowers this minimum, thereby enhancing the biphoton purity.

\figfour

\subsection{Enhanced Impact of Higher Optical Depth}

Figure~\ref{fig4}(a) shows the coincidence count rates measured at an OD of 35. In previous measurements, a total accumulation time of 1 hour was sufficient to clearly reveal the biphoton wavepacket. However, the accumulation of background counts during this period was inadequate, resulting in an incomplete background baseline. As a consequence, the zero-delay conditional autocorrelation function failed to exhibit the expected value of 2 for thermal background. To address this issue, we extended the coincidence accumulation time to 6 hours for a more reliable comparison. The experimental results yield $R_{s}^{\rm exp}=8.9(2)\times10^5$ s$^{-1}$, $R_{as}^{\rm exp}=8.5(2)\times10^5$ s$^{-1}$, $r_{p,s}^{\rm exp}=0.79(1)$, and $r_{p,as}^{\rm exp}=0.83(1)$. These uncertainties correspond to the standard deviations obtained from six independent 1-hour experimental runs. Here, the pairing ratios are larger than those obtained at ${\rm OD}=20$. This increase is attributed to the higher likelihood of photons interacting with other atoms as they propagate through the ensemble at higher OD, thereby enhancing the efficiency of stimulated FWM. An increase in OD also extends the $\tau_c$ of the unconditional autocorrelation function $g_{as\text{-}as}^{(2)}$ to approximately $195$ ns, as shown in the inset. This phenomenon arises from the built-in $\Lambda$-type EIT acting as a frequency filter: a higher OD narrows the transparency window, reducing the photon bandwidth and thereby increasing $\tau_c$. The extended $\tau_c$ benefits quantum communication by enabling photons to maintain phase coherence over longer distances. Furthermore, this effect also increases the biphoton coherence time, with $\tau_{\rm delay} \approx 134$ ns compared to $81$ ns in Fig.~\ref{fig2}(b) for ${\rm OD}=20$.

Figure~\ref{fig4}(b) shows the coincidence count rates and $g_{as\text{-}as}^{(2)}(\tau)$ for the case of $\gamma_{21}=0.1\Gamma$. The higher OD relative to ${\rm OD}=20$ makes this setup more sensitive to the effects of $\gamma_{21}$. In this case, the coherence times $\tau_{\rm delay}$ and $\tau_c$ decrease to $107$ ns and $154$ ns, respectively. These reductions enhance $r_{\rm SB}$ from 9.7(2) at $\gamma_{21}=0.001\Gamma$ to 11.7(3) at $\gamma_{21}=0.1\Gamma$. Furthermore, the asymmetry in $R_{{\rm C},s}$ and $R_{{\rm C},as}$ becomes more pronounced at higher OD. At ${\rm OD}=35$, the measured generation rates are $R_{s}^{\rm exp}=8.7(2)\times10^5$ $\mathrm{s}^{-1}$ and $R_{as}^{\rm exp}=6.0(2)\times10^5$ $\mathrm{s}^{-1}$. The discrepancy $(R_{s}^{\rm exp}-R_{as}^{\rm exp})/R_{s}^{\rm exp}$ of 31\% is larger than the 21\% observed when ${\rm OD} = 20$. This occurs because, at the same $\gamma_{21}$, a higher OD leads to greater dissipation of anti-Stokes photons, thereby accentuating asymmetric biphoton generation. The difference in correlated areas directly represents the pairing ratios, with measured values of $r_{p,s}^{\rm exp}=0.55(2)$ and $r_{p,as}^{\rm exp}=0.79(2)$. The discrepancy $(r_{p,as}^{\rm exp}-r_{p,s}^{\rm exp})/r_{p,as}^{\rm exp}=31\%$ at ${\rm OD}=35$ is larger than the $21\%$ observed at ${\rm OD}=20$, indicating that higher OD amplifies the difference between $r_{p,s}$ and $r_{p,as}$ under the same $\gamma_{21}$. 

Such asymmetric biphoton generation directly influences the heralding efficiency $\eta_h$, which serves as a key performance metric that depends on both the intrinsic properties of the biphoton source and the collection efficiencies in the Stokes and anti-Stokes channels, $\eta_s$ and $\eta_{as}$. Consider the case of symmetric biphoton generation, where $R_s = R_{as} = R$ and $r_{p,s} = r_{p,as} = r_p$. If Stokes photons are chosen as the trigger, the heralding efficiency $\eta_{h,s}$ is defined as the fraction of received anti-Stokes photons that are heralded. This is obtained by dividing the detection rate of heralded single photons, $D_h = R \eta_s \eta_{as} r_p$, by the detection rate of anti-Stokes photons, $D_{as} = R \eta_{as}$, which yields $\eta_{h,s} = \eta_s r_p$. Conversely, if anti-Stokes photons serve as the trigger, the heralding efficiency becomes $\eta_{h,as} = \eta_{as} r_p$. Therefore, in the symmetric case, selecting the channel with the higher collection efficiency as the heralding channel enhances the overall heralding efficiency and facilitates further quantum processing tasks. However, when asymmetric biphoton generation occurs, the expressions change to $\eta_{h,s} = \eta_s r_{p,as}$ and $\eta_{h,as} = \eta_{as} r_{p,s}$. This indicates that, in addition to the collection efficiency, the asymmetric pairing ratio must be carefully considered to optimize the heralding efficiency.

Figures \ref{fig4}(c) and \ref{fig4}(d) show $g_{as\text{-}as|s}^{(2)}(\tau)$ under $\gamma_{21}=0.001\Gamma$ and $\gamma_{21}=0.1\Gamma$, respectively. The experimentally obtained  minima are 0.37(1) and 0.31(1), as a clear indication that ground-state decoherence enhances biphoton purity. It is evident that under the condition of ${\rm OD}=35$, the difference in the minimum $g_{as\text{-}as|s}^{(2)}(\tau)$ between $\gamma_{21}=0.1\Gamma$ and $\gamma_{21}=0.001\Gamma$ is more pronounced than that observed at ${\rm OD}=20$. This is because $\gamma_{21}$ imposes an upper limit on $\tau_{\rm delay}$~\cite{SFWM8}. In other words, when $\tau_{\rm delay}$ is relatively short, a larger $\gamma_{21}$ is required to induce a noticeable effect. This explains why, despite increasing $\gamma_{21}$ to $0.1\Gamma$, its impact on the biphoton wavepacket remains limited. In contrast, since ${\rm OD}=35$ extends $\tau_{\rm delay}$, the biphoton wavepacket becomes more susceptible to the influence of $\gamma_{21}=0.1\Gamma$, leading to a more significant increase in $g_{as\text{-}as|s}^{(2)}(\tau)$.

Apart from the difference in the minimum values of $g_{as\text{-}as|s}^{(2)}(\tau)$, the background at ${\rm OD}=35$ appears smoother and more stable, owing to the extended data acquisition time and the higher photon generation rate. However, we emphasize that in practical applications, it is not necessary to accumulate data until the background becomes completely smooth, as these contributions originate from uncorrelated photons following a thermal distribution and are typically irrelevant to quantum information tasks. It is also worth noting that, although the generation rate at ${\rm OD} = 35$ is significantly higher than that at ${\rm OD} = 20$, the quantum-state purity of the heralded anti-Stokes photons is reduced due to the increased contribution from uncorrelated background photons. This observation underscores a fundamental trade-off between photon generation rate and heralded single-photon purity, where neither parameter is universally optimal; the appropriate balance depends on the specific requirements of the quantum system in which the biphotons are utilized.


\section{Conclusion}

In this study, we present an experimental investigation of asymmetric biphoton generation in a double-$\Lambda$ SFWM scheme, highlighting significant differences in the Stokes and anti-Stokes photon generation rates. The observed asymmetry arises from the interplay of phase mismatch and ground-state decoherence, both of which degrade the generation of temporally correlated photons. Under minimal ground-state decoherence, phase mismatch inherent in the backward configuration primarily drives the asymmetry. As ground-state decoherence increases, it further amplifies the asymmetry, underscoring its pivotal role in shaping biphoton properties. Our findings challenge the conventional interpretation of biphotons as a simple two-mode squeezed state, demonstrating that the stimulated FWM process inherent in SFWM provides a more accurate characterization. Using the coincidence count rate representation, we offer deeper insights into biphoton sources, including photon generation rates and pairing ratios, extending beyond standard analyses. Furthermore, we establish a direct conversion between the cross-correlation and conditional autocorrelation functions, revealing that ground-state decoherence, while reducing temporally correlated photons, paradoxically enhances biphoton purity—a counterintuitive phenomenon reported here for the first time. Additionally, our unconditional autocorrelation measurements confirm that the generated photons follow a thermal-state distribution, consistent with theoretical predictions. By comprehensively analyzing biphoton generation dynamics, this work advances the understanding of SFWM processes and provides valuable insights for optimizing biphoton sources. These findings hold significant implications for quantum communication, quantum metrology, and other quantum technologies reliant on high-purity, asymmetric biphoton sources.


\section*{Appendix A: Explicit Expressions of Theoretical Parameters}
The Stokes field operator $\widetilde{a}_{s}$ and the anti-Stokes field operator $\widetilde{a}_{as}$ evolve according to the Maxwell–Schrödinger equations (MSEs), as shown in Eqs.~\eqref{eq1} and \eqref{eq2}, where the light-matter interaction is encoded in the atomic response operators $\widetilde{\sigma}_{jk}$. These operators describe the collective response of the medium and serve as the key link between the optical fields and the atomic ensemble. By solving the corresponding Heisenberg–Langevin equations (HLEs), the atomic coherence associated with the Stokes and anti-Stokes fields can be expressed in the following explicit form:
\begin{align}
\widetilde{\sigma}_{23(41)}
=\epsilon_{23(41)}g_s\widetilde{a}_s
 +\eta_{23(41)}g_{as}^*\widetilde{a}_{as}^\dagger
 +\sum_{jk}\zeta_{jk}^{s(as)}\widetilde{F}_{jk}
=\frac{\bar{\epsilon}_{23(41)}}{T}g_s\widetilde{a}_s
 +\frac{\bar{\eta}_{23(41)}}{T}g_{as}^*\widetilde{a}_{as}^\dagger 
 +\sum_{jk}\frac{\bar{\zeta}_{jk}^{s(as)}}{T}\widetilde{F}_{jk}
,\tag{A1}\label{A1}
\end{align} 
where $T=(|\Omega_c|^2-|\Omega_d|^2)^2+|\Omega_c|^2(\bar{\gamma}_{21}\bar{\gamma}_{41}+\bar{\gamma}_{23}\bar{\gamma}_{43})+|\Omega_d|^2(\bar{\gamma}_{21}\bar{\gamma}_{23}+\bar{\gamma}_{41}\bar{\gamma}_{43})+\bar{\gamma}_{21}\bar{\gamma}_{23}\bar{\gamma}_{41}\bar{\gamma}_{43}$ with the definitions $\bar{\gamma}_{21}=\gamma_{21}-2i\omega$, $\bar{\gamma}_{23}=\Gamma-2i\Delta_d-2i\omega$, $\bar{\gamma}_{41}=\Gamma+2i\Delta_c-2i\omega$, and $\bar{\gamma}_{43}=2\Gamma-2i\Delta_d+2i\Delta_c-2i\omega$. Here, $\Delta_c=\omega_c-\omega_{42}$ is the coupling detuning. The explicit forms of $\bar{\epsilon}_{23(41)}$, $\bar{\eta}_{23(41)}$, and $\bar{\zeta}_{jk}^{s(as)}$ are given by
\begin{align}
\bar{\epsilon}_{23}=&
2i
(\langle\hat{\sigma}_{22}\rangle-\langle\hat{\sigma}_{33}\rangle)
(\bar{\gamma}_{21}\bar{\gamma}_{41}\bar{\gamma}_{43}
+\bar{\gamma}_{43}|\Omega_c|^2
+\bar{\gamma}_{21}|\Omega_d|^2)
+
2\Omega_d\langle\hat{\sigma}_{31}\rangle
(\bar{\gamma}_{41}\bar{\gamma}_{43}-|\Omega_c|^2+|\Omega_d|^2)
\nonumber\\
&+
2\Omega_c\langle\hat{\sigma}_{42}\rangle
(\bar{\gamma}_{21}\bar{\gamma}_{41}+|\Omega_c|^2-|\Omega_d|^2)
,\tag{A2a}\label{A2a}
\\
\bar{\eta}_{23}=&
-2i\Omega_d\Omega_c
(\langle\hat{\sigma}_{11}\rangle-\langle\hat{\sigma}_{44}\rangle)
(\bar{\gamma}_{21}+\bar{\gamma}_{43})
-
2\Omega_c\langle\hat{\sigma}_{13}\rangle
(\bar{\gamma}_{21}\bar{\gamma}_{41}+|\Omega_c|^2-|\Omega_d|^2)
\nonumber\\
&-
2\Omega_d\langle\hat{\sigma}_{24}\rangle
(\bar{\gamma}_{41}\bar{\gamma}_{43}-|\Omega_c|^2+|\Omega_d|^2)
,\tag{A2b}\label{A2b}
\\
\bar{\epsilon}_{41}=&
2i\Omega_d^*\Omega_c^*
(\langle\hat{\sigma}_{22}\rangle-\langle\hat{\sigma}_{33}\rangle)
(\bar{\gamma}_{21}+\bar{\gamma}_{43})
-
2\Omega_c^*\langle\hat{\sigma}_{31}\rangle
(\bar{\gamma}_{23}\bar{\gamma}_{43}+|\Omega_c|^2-|\Omega_d|^2)
\nonumber\\
&-
2\Omega_d^*\langle\hat{\sigma}_{42}\rangle
(\bar{\gamma}_{21}\bar{\gamma}_{23}-|\Omega_c|^2+|\Omega_d|^2)
,\tag{A2c}\label{A2c}
\\
\bar{\eta}_{41}=&
-2i
(\langle\hat{\sigma}_{11}\rangle-\langle\hat{\sigma}_{44}\rangle)
(\bar{\gamma}_{21}\bar{\gamma}_{23}\bar{\gamma}_{43}+\bar{\gamma}_{21}|\Omega_c|^2+\bar{\gamma}_{43}|\Omega_d|^2)
+
2\Omega_d^*\langle\hat{\sigma}_{13}\rangle
(\bar{\gamma}_{21}\bar{\gamma}_{23}-|\Omega_c|^2+|\Omega_d|^2)\nonumber\\
&+
2\Omega_c^*\langle\hat{\sigma}_{24}\rangle
(\bar{\gamma}_{23}\bar{\gamma}_{43}+|\Omega_c|^2-|\Omega_d|^2)
,\tag{A2d}\label{A2d}
\end{align}
\begin{align}
\bar{\zeta}_{21}^{s}
=&
2i\Omega_d
(
 \bar{\gamma}_{41}\bar{\gamma}_{43}
 -|\Omega_c|^2+|\Omega_d|^2
)
,\tag{A2e}\label{A2e}
\\
\bar{\zeta}_{23}^{s}
=&
2
(
 \bar{\gamma}_{21}\bar{\gamma}_{41}\bar{\gamma}_{43}
 +\bar{\gamma}_{43}|\Omega_c|^2
 +\bar{\gamma}_{21}|\Omega_d|^2
)
,\tag{A2f}\label{A2f}
\\
\bar{\zeta}_{41}^{s}
=&
2\Omega_d\Omega_c
(
 \bar{\gamma}_{21}+\bar{\gamma}_{43}
)
,\tag{A2g}\label{A2g}
\\
\bar{\zeta}_{43}^{s}
=&
-2i\Omega_c
(
 \bar{\gamma}_{21}\bar{\gamma}_{41}
 +|\Omega_c|^2
 -|\Omega_d|^2
)
,\tag{A2h}\label{A2h}
\\
\bar{\zeta}_{21}^{as}
=&
-2i\Omega_c^*
(
 \bar{\gamma}_{23}\bar{\gamma}_{43}
 +|\Omega_c|^2-|\Omega_d|^2
)
,\tag{A2i}\label{A2i}
\\
\bar{\zeta}_{23}^{as}
=&
2\Omega_d^*\Omega_c^*
(
 \bar{\gamma}_{21}+\bar{\gamma}_{43}
)
,\tag{A2j}\label{A2j}
\\
\bar{\zeta}_{41}^{as}
=&
2
(
 \bar{\gamma}_{21}\bar{\gamma}_{23}\bar{\gamma}_{43}
 +\bar{\gamma}_{21}|\Omega_c|^2
 +\bar{\gamma}_{43}|\Omega_d|^2
)
,\tag{A2k}\label{A2k}
\\
\bar{\zeta}_{43}^{as}
=&
2i\Omega_d^*
(
 \bar{\gamma}_{21}\bar{\gamma}_{23}
 -|\Omega_c|^2
 +|\Omega_d|^2
)
.\tag{A2l}\label{A2l}
\end{align}
Several expectation values of $\hat{\sigma}_{jk}$ can be obtained by solving the zeroth-order HLEs:
$\langle\hat{\sigma}_{11}\rangle=\frac{|\Omega_c|^2(\Gamma^2+4\Delta_d^2)+|\Omega_d|^2|\Omega_c|^2}{M}$, 
$\langle\hat{\sigma}_{22}\rangle=\frac{|\Omega_d|^2(\Gamma^2+4\Delta_c^2)+|\Omega_d|^2|\Omega_c|^2}{M}$, 
$\langle\hat{\sigma}_{33}\rangle=\langle\hat{\sigma}_{44}\rangle=\frac{|\Omega_d|^2|\Omega_c|^2}{M}$, 
$\langle\hat{\sigma}_{13}\rangle=\frac{i(\Gamma+2i\Delta_d)|\Omega_c|^2\Omega_d}{M}$, and 
$\langle\hat{\sigma}_{24}\rangle=\frac{i(\Gamma+2i\Delta_c)|\Omega_d|^2\Omega_c}{M}$,
where $M=|\Omega_d|^2(\Gamma^2+4\Delta_c^2)+|\Omega_c|^2(\Gamma^2+4\Delta_d^2)+4|\Omega_d|^2|\Omega_c|^2$. 
By substituting these expressions, MSEs can be further rewritten as follows:
\begin{align}
\frac{\partial}{\partial z}
\begin{bmatrix}
\widetilde{a}_s \\ \widetilde{a}_{as}^\dagger
\end{bmatrix}
=&
\begin{bmatrix}
g_R & \kappa_s \\ \kappa_{as} & \Gamma_{as}+i\Delta k
\end{bmatrix}
\begin{bmatrix}
\widetilde{a}_s \\ \widetilde{a}_{as}^\dagger
\end{bmatrix}
+\sqrt{\frac{N}{c}}\sum_{jk}
\begin{bmatrix}
\xi_{jk}^s \\ \xi_{jk}^{as}
\end{bmatrix}
\widetilde{F}_{jk}
.\tag{A3}\label{A3}
\end{align}
The Raman gain coefficient $g_R = \frac{iN|g_s|^2}{c}\epsilon_{23} + \frac{i\omega}{c}$ characterizes the self-susceptibility of the Stokes photons and describes the Raman process involved in biphoton generation. The coefficient $\Gamma_{as} = \frac{iN|g_s|^2}{c}\eta_{41} - \frac{i\omega}{c}$ corresponds to the self-susceptibility of the anti-Stokes photons and determines the EIT spectral profile. When $\Delta_d$ is large, this coefficient exhibits the dispersive behavior associated with $\Lambda$-type EIT. The cross-susceptibility is described by the coupling coefficients $\kappa_s=\frac{iN g_s^* g_{as}^*}{c}\eta_{23}$ and $\kappa_{as} = \frac{iN g_s g_{as}}{c} \epsilon_{41}$, which quantify how the anti-Stokes (Stokes) photons influence the Stokes (anti-Stokes) field through their mutual coupling. The vacuum fluctuation contributions are captured by the coefficients $\xi_{jk}^s = i g_s^* \sqrt{\frac{N}{c}} \zeta_{jk}^s$ and $\xi_{jk}^{as} = i g_{as} \sqrt{\frac{N}{c}} \zeta_{jk}^{as}$.

In our experimental system, the phases of $g_s$ and $g_{as}$ can be safely neglected, as they do not affect the calculation of generation rates or correlation functions. Furthermore, since $g_s = g_{as} \equiv g$ in our configuration, the quantity $\frac{g^2 N}{c}$ can be replaced by the experimentally relevant expression $\frac{{\rm OD} \Gamma }{4L}$, where $\Gamma$ is the spontaneous decay rate of rubidium-87 atoms and $L$ is the length of the atomic ensemble. By integrating over the propagation distance from $z = 0$ to $z = L$, we obtain:
\begin{align}
&\begin{bmatrix}
\widetilde{a}_s(L,\omega) \\ \widetilde{a}_{as}^\dagger(L,-\omega)
\end{bmatrix}
=
\begin{bmatrix}
A'(\omega) & B'(\omega) \\ C'(\omega) & D'(\omega)
\end{bmatrix}
\begin{bmatrix}
\widetilde{a}_s(0,\omega) \\ \widetilde{a}_{as}^\dagger(0,-\omega)
\end{bmatrix}
+\sqrt{\frac{N}{c}}
\sum_{jk}
\int_0^Ldz\,
\begin{bmatrix}
P'_{jk}(z,\omega) \\ Q'_{jk}(z,\omega)
\end{bmatrix}\widetilde{F}_{jk}(z,\omega)
.\tag{A4}\label{A4}
\end{align}
where $\begin{bmatrix}\begin{smallmatrix}
A'(\omega) & B'(\omega) \\ C'(\omega) & D'(\omega)
\end{smallmatrix}\end{bmatrix}=\exp \left(
\begin{bmatrix}\begin{smallmatrix}
g_R(\omega) & \kappa_s(\omega) \\ \kappa_{as}(\omega) & \Gamma_{as}(\omega)+i\Delta k
\end{smallmatrix}\end{bmatrix}L
\right)$ and 
$\begin{bmatrix}\begin{smallmatrix}
P'_{jk}(z,\omega) \\ Q'_{jk}(z,\omega)
\end{smallmatrix}\end{bmatrix}=
\exp\left(
\begin{bmatrix}\begin{smallmatrix}
g_R(\omega) & \kappa_s(\omega) \\ \kappa_{as}(\omega) & \Gamma_{as}(\omega)+i\Delta k
\end{smallmatrix}\end{bmatrix}(L-z)
\right)
\begin{bmatrix}\begin{smallmatrix}
\xi_{jk}^s(\omega) \\ \xi_{jk}^{as}(\omega)
\end{smallmatrix}\end{bmatrix}$.
We rearrange Eq. \eqref{A4} such that the output field operators, $\widetilde{a}_{s}(L,\omega)$ and $\widetilde{a}_{as}^{\dagger}(0,-\omega)$, are expressed as linear superpositions of the input operators. This transformation yields the form presented in Eq.~\eqref{eq3}, where the new coefficients are defined as $A(\omega)=A'(\omega)-\frac{B'(\omega)C'(\omega)}{D'(\omega)}$, $B(\omega)=\frac{B'(\omega)}{D'(\omega)}$, $C(\omega)=-\frac{C'(\omega)}{D'(\omega)}$, $D(\omega)=\frac{1}{D'(\omega)}$, $P_{jk}(z,\omega)=P_{jk}'(z,\omega)-\frac{B'(\omega)}{D'(\omega)}Q_{jk}'(z,\omega)$, and $Q_{jk}(z,\omega)=-\frac{1}{D'(\omega)}Q_{jk}'(z,\omega)$. In addition, the associated diffusion coefficients can be derived using the Einstein relation $\mathcal{D}_{jk,j'k'}=\frac{\partial}{\partial t}\langle\hat{\sigma}_{jk}\hat{\sigma}_{j'k'}\rangle-\langle\hat{R}_{jk}\hat{\sigma}_{j'k'}\rangle-\langle\hat{\sigma}_{jk}\hat{R}_{j'k'}\rangle$~\cite{app8,Shiu1,Scully}, which yields the following expressions:
\begin{align}
&\begin{bmatrix}\begin{smallmatrix}
\mathcal{D}_{12,21} & \mathcal{D}_{12,23} & \mathcal{D}_{12,41} & \mathcal{D}_{12,43} \vphantom{\frac{\gamma_{21}}{2}} \\
\mathcal{D}_{32,21} & \mathcal{D}_{32,23} & \mathcal{D}_{32,41} & \mathcal{D}_{32,43} \vphantom{\frac{\gamma_{21}}{2}} \\
\mathcal{D}_{14,21} & \mathcal{D}_{14,23} & \mathcal{D}_{14,41} & \mathcal{D}_{14,43} \vphantom{\frac{\gamma_{21}}{2}} \\
\mathcal{D}_{34,21} & \mathcal{D}_{34,23} & \mathcal{D}_{34,41} & \mathcal{D}_{34,43} \vphantom{\frac{\gamma_{21}}{2}}
\end{smallmatrix}\end{bmatrix}
=
\begin{bmatrix}\begin{smallmatrix}
\gamma_{21}\langle\hat{\sigma}_{11}\rangle + \frac{\Gamma}{2}\langle\hat{\sigma}_{33}\rangle + \frac{\Gamma}{2}\langle\hat{\sigma}_{44}\rangle & \frac{\gamma_{21}}{2}\langle\hat{\sigma}_{13}\rangle & 0 & 0
\\
\frac{\gamma_{21}}{2}\langle\hat{\sigma}_{31}\rangle & 0 & 0 & 0
\\
0 & 0 & \Gamma\langle\hat{\sigma}_{11}\rangle + \frac{\Gamma}{2}\langle\hat{\sigma}_{33}\rangle + \frac{\Gamma}{2}\langle\hat{\sigma}_{44}\rangle & \Gamma\langle\hat{\sigma}_{13}\rangle
\\
0 & 0 & \Gamma\langle\hat{\sigma}_{31}\rangle & \Gamma\langle\hat{\sigma}_{33}\rangle
\end{smallmatrix}\end{bmatrix}
,\tag{A5a}\label{A5a}
\\
&\begin{bmatrix}\begin{smallmatrix}
\mathcal{D}_{21,12} & \mathcal{D}_{21,32} & \mathcal{D}_{21,14} & \mathcal{D}_{21,34} \vphantom{\frac{\gamma_{21}}{2}} \\
\mathcal{D}_{23,12} & \mathcal{D}_{23,32} & \mathcal{D}_{23,14} & \mathcal{D}_{23,34} \vphantom{\frac{\gamma_{21}}{2}} \\
\mathcal{D}_{41,12} & \mathcal{D}_{41,32} & \mathcal{D}_{41,14} & \mathcal{D}_{41,34} \vphantom{\frac{\gamma_{21}}{2}} \\
\mathcal{D}_{43,12} & \mathcal{D}_{43,32} & \mathcal{D}_{43,14} & \mathcal{D}_{43,34} \vphantom{\frac{\gamma_{21}}{2}}
\end{smallmatrix}\end{bmatrix}
=
\begin{bmatrix}\begin{smallmatrix}
\gamma_{21}\langle\hat{\sigma}_{22}\rangle + \frac{\Gamma}{2}\langle\hat{\sigma}_{33}\rangle + \frac{\Gamma}{2}\langle\hat{\sigma}_{44}\rangle & 0 & \frac{\gamma_{21}}{2}\langle\hat{\sigma}_{24}\rangle & 0
\\
0 & \Gamma\langle\hat{\sigma}_{22}\rangle + \frac{\Gamma}{2}\langle\hat{\sigma}_{33}\rangle + \frac{\Gamma}{2}\langle\hat{\sigma}_{44}\rangle & 0 & \Gamma\langle\hat{\sigma}_{24}\rangle
\\
\frac{\gamma_{21}}{2}\langle\hat{\sigma}_{42}\rangle & 0 & 0 & 0
\\
0 & \Gamma\langle\hat{\sigma}_{42}\rangle & 0 & \Gamma\langle\hat{\sigma}_{44}\rangle
\end{smallmatrix}\end{bmatrix}
.\tag{A5b}\label{A5b}
\end{align}
It is important to note that the subscripts of the diffusion coefficients in Eqs.~\eqref{eq4} and \eqref{eq5} differ from the notation used here. Specifically, they obey the relations $\mathcal{D}_{jk^\dagger,j'k'} = \mathcal{D}_{kj,j'k'}$ and $\mathcal{D}_{jk,j'k'^\dagger} = \mathcal{D}_{jk,k'j'}$, respectively.


\section*{Appendix B: Derivation of the Conditional Autocorrelation Function}

The conditional autocorrelation function is defined as below:
\begin{align}
g_{as\text{-}as|s}^{(2)}(t,\tau_1,\tau_2)=
\frac{\langle\hat{a}_s^\dagger(t)\hat{a}_s(t)\rangle
\langle\hat{a}_s^\dagger(t)\hat{a}_{as,1}^\dagger(t+\tau_1)\hat{a}_{as,2}^\dagger(t+\tau_2)\hat{a}_{as,2}(t+\tau_2)\hat{a}_{as,1}(t+\tau_1)\hat{a}_s(t)\rangle}
{\langle\hat{a}_s^\dagger(t)\hat{a}_{as,1}^\dagger(t+\tau_1)\hat{a}_{as,1}(t+\tau_1)\hat{a}_s(t)\rangle\langle\hat{a}_s^\dagger(t)\hat{a}_{as,2}^\dagger(t+\tau_2)\hat{a}_{as,2}(t+\tau_2)\hat{a}_s(t)\rangle}
.
\tag{B1}
\label{B1}
\end{align}
This equation describes the autocorrelation function of anti-Stokes photons measured after detecting Stokes photons. Here, $\tau_1$ and $\tau_2$ represent the time differences between the detection of two anti-Stokes channels and the Stokes photons. In this work, since we focus on the zero-delay conditional autocorrelation function, we set $\tau_1 = \tau_2=\tau$. To analyze the relationship between $\hat{a}_{as,1}(t+\tau)$, $\hat{a}_{as,2}(t+\tau)$, and $\hat{a}_{s}(t)$, we consider the operators before and after the beam splitter, which can be expressed as follows:
\begin{align}
\mathcal{E}_{as}\hat{a}_{as}+{\rm h.c.}
\rightarrow k_1\mathcal{E}_{as}\hat{a}_{as,1}+k_2\mathcal{E}_{as}\hat{a}_{as,2}+{\rm h.c.},
\tag{B2}
\label{B2}
\end{align}
where $|k_1|^2 = |k_2|^2 = 1/2$ for a perfect beam splitter. Since the anti-Stokes field passing through the beam splitter does not change the characteristics of the field operators, we have $\hat{a}_{as,1}=\frac{e^{i\phi_1}}{\sqrt{2}}\hat{a}_{as}$ and $\hat{a}_{as,2}=\frac{e^{i\phi_2}}{\sqrt{2}}\hat{a}_{as}$, where $\phi_1$ and $\phi_2$ are phases arising from the paths. Utilizing the generalized Wick's theorem, the zero-delay conditional autocorrelation function can be derived as follows:
\begin{align}
g_{as\text{-}as|s}^{(2)}(t,\tau,\tau)
=&
\frac{\langle\hat{a}_s^\dagger\hat{a}_s\rangle
\langle\hat{a}_s^\dagger\hat{a}_{as}^\dagger\hat{a}_{as}^\dagger
\hat{a}_{as}\hat{a}_{as}\hat{a}_s\rangle}
{\langle\hat{a}_s^\dagger\hat{a}_{as}^\dagger\hat{a}_{as}\hat{a}_s\rangle\langle\hat{a}_s^\dagger\hat{a}_{as}^\dagger\hat{a}_{as}\hat{a}_s\rangle}
\nonumber\\
=&
\langle\hat{a}_s^\dagger\hat{a}_s\rangle\frac
{
4\langle\hat{a}_s^\dagger\hat{a}_{as}^\dagger\rangle
\langle\hat{a}_{as}^\dagger\hat{a}_{as}\rangle\langle\hat{a}_{as}\hat{a}_s\rangle
+
2\langle\hat{a}_s^\dagger\hat{a}_s\rangle
\langle\hat{a}_{as}^\dagger\hat{a}_{as}\rangle\langle
\hat{a}_{as}^\dagger\hat{a}_{as}\rangle
}
{\langle\hat{a}_s^\dagger\hat{a}_{as}^\dagger\hat{a}_{as}\hat{a}_s\rangle\langle\hat{a}_s^\dagger\hat{a}_{as}^\dagger\hat{a}_{as}\hat{a}_s\rangle}
\nonumber\\
=&
\frac{1}{\langle\hat{a}_s^\dagger\hat{a}_s\rangle\langle\hat{a}_{as}^\dagger\hat{a}_{as}\rangle}\frac
{
4\left(\langle\hat{a}_s^\dagger\hat{a}_{as}^\dagger\rangle
\langle\hat{a}_{as}\hat{a}_s\rangle
+\langle\hat{a}_s^\dagger\hat{a}_s\rangle
\langle\hat{a}_{as}^\dagger\hat{a}_{as}\rangle\right)
-2\langle\hat{a}_s^\dagger\hat{a}_s\rangle
\langle\hat{a}_{as}^\dagger\hat{a}_{as}\rangle
}
{\left[g_{s\text{-}as}^{(2)}(\tau)\right]^2}
=
\frac
{4g_{s\text{-}as}^{(2)}(\tau)-2}
{\left[g_{s\text{-}as}^{(2)}(\tau)\right]^2}.
\tag{B3}
\label{B3}
\end{align}
For notation simplicity, we replace $g_{as\text{-}as|s}^{(2)}(t,\tau,\tau)$ with $g_{as\text{-}as|s}^{(2)}(\tau)$.

\section*{Acknowledgements}

This work was supported by the National Science and Technology Council of Taiwan under Grants No. 112-2112-M-006-034, No. 113-2112-M-006-028, and No. 113-2119-M-007-012. We also acknowledge support from the Center for Quantum Science and Technology (CQST) within the framework of the Higher Education Sprout Project by the Ministry of Education (MOE) in Taiwan.

\section*{Conflict of Interest}

The authors declare no conflict of interest.

\section*{Data Availability Statement}

The data that support the ﬁndings of this study are available from the corresponding author upon reasonable request.

\section*{Keywords}

biphotons, spontaneous four-wave mixing, pairing ratio



\begin{thebibliography}{99}
	
\bibitem{QKD1}
P. W. Shor, J. Preskill,
\textit{Phys. Rev. Lett.} \textbf{2000}, 85, 441.

\bibitem{QKD2}
V. Scarani, H. Bechmann-Pasquinucci, N. J. Cerf, M. Du\v{s}ek, N. L\"utkenhaus, M. Peev,
\textit{Rev. Mod. Phys.} \textbf{2009}, 81, 1301.

\bibitem{QKD3}
F. Xu, X. Ma, Q. Zhang, H.-K. Lo, J.-W. Pan,
\textit{Rev. Mod. Phys.} \textbf{2020}, 92, 025002.

\bibitem{QKD4}
Y. Liu, W.-J. Zhang, C. Jiang, J.-P. Chen, C. Zhang, W.-X. Pan, D. Ma, H. Dong, J.-M. Xiong, C.-J. Zhang, H. Li, R.-C. Wang, J. Wu, T.-Y. Chen, L. You, X.-B. Wang, Q. Zhang, J.-W. Pan,
\textit{Phys. Rev. Lett.} \textbf{2023}, 130, 210801.


\bibitem{comp1}
P. Kok, W. J. Munro, K. Nemoto, T. C. Ralph, J. P. Dowling, G. J. Milburn,
\textit{Rev. Mod. Phys.} \textbf{2007}, 79, 135.

\bibitem{comp2}
J. L. O'Brien,
\textit{Science} \textbf{2007}, 318, 1567.

\bibitem{comp3}
H.-S. Zhong, H. Wang, Y.-H. Deng, M.-C. Chen, L.-C. Peng, Y.-H. Luo, J. Qin, D. Wu, X. Ding, Y. Hu, P. Hu, X.-Y. Yang, W.-J. Zhang, H. Li, Y. Li, X. Jiang, L. Gan, G. Yang, L. You, Z. Wang, L. Li, N.-L. Liu, C.-Y. Lu, J.-W. Pan,
\textit{Science} \textbf{2020}, 370, 1460.


\bibitem{tele1}
D. Bouwmeester, J.-W. Pan, K. Mattle, M. Eibl, H. Weinfurter, A. Zeilinger,
\textit{Nature} \textbf{1997}, 390, 575.

\bibitem{tele2}
J. F. Sherson, H. Krauter, R. K. Olsson, B. Julsgaard, K. Hammerer, I. Cirac, E. S. Polzik,
\textit{Nature} \textbf{2006}, 443, 557.

\bibitem{tele3}
J.-G. Ren, P. Xu, H.-L. Yong, L. Zhang, S.-K. Liao, J. Yin, W.-Y. Liu, W.-Q. Cai, M. Yang, L. Li, K.-X. Yang, X. Han, Y.-Q. Yao, J. Li, H.-Y. Wu, S. Wan, L. Liu, D.-Q. Liu, Y.-W. Kuang, Z.-P. He, P. Shang, C. Guo, R.-H. Zheng, K. Tian, Z.-C. Zhu, N.-L. Liu, C.-Y. Lu, R. Shu, Y.-A. Chen, C.-Z. Peng, J.-Y. Wang, J.-W. Pan,
\textit{Nature} \textbf{2017}, 549, 70.



\bibitem{FT1}
Y.-W. Cho, K.-K. Park, J.-C. Lee, Y.-H. Kim,
\textit{Phys. Rev. Lett.} \textbf{2014}, 113, 063602.

\bibitem{FT2}
J. Park, T. Jeong, H. Kim, H. S. Moon,
\textit{Phys. Rev. Lett.} \textbf{2018}, 121, 263601.

\bibitem{FT3}
G.-H. Lee, Y. S. Ihn, A. Lee, U.-S. Kim, Y.-H. Kim,
\textit{Phys. Rev. A} \textbf{2019}, 100, 053817.

\bibitem{FT4}
Y. Mei, Y. Zhou, S. Zhang, J. Li, K. Liao, H. Yan, S.-L. Zhu, S. Du,
\textit{Phys. Rev. Lett.} \textbf{2020}, 124, 010509.


\bibitem{pol1}
H. Yan, S. Zhang, J. F. Chen, M. M. T. Loy, G. K. L. Wong, S. Du, 
\textit{Phys. Rev. Lett.} \textbf{2011}, 106, 033601. 

\bibitem{pol2}
K. Liao, H. Yan, J. He, S. Du, Z.-M. Zhang, S.-L. Zhu, 
\textit{Phys. Rev. Lett.} \textbf{2014}, 112, 243602. 

\bibitem{pol3}
Y. Wang, J. Li, S. Zhang, K. Su, Y. Zhou, K. Liao, S. Du, H. Yan, S.-L. Zhu, 
\textit{Nat. Photonics} \textbf{2019}, 13, 346. 

\bibitem{pol4}
J. Park, H. Kim, H. S. Moon, 
\textit{Phys. Rev. Lett.} \textbf{2019}, 122, 143601. 


\bibitem{OAM1}
Q.-F. Chen, B.-S. Shi, Y.-S. Zhang, G.-C. Guo, 
\textit{Phys. Rev. A} \textbf{2008}, 78, 053810. 

\bibitem{OAM2}
D.-S. Ding, Z.-Y. Zhou, B.-S. Shi, G.-C. Guo, 
\textit{Nat. Commun.} \textbf{2013}, 4, 2527. 

\bibitem{OAM3}
T.-M. Zhao, Y. S. Ihn, Y.-H. Kim, 
\textit{Phys. Rev. Lett.} \textbf{2019}, 122, 123607. 



\bibitem{SPDC1}
P.-J. Tsai, Y.-C. Chen, 
\textit{Quantum Sci. Technol.} \textbf{2018}, 3, 034005. 

\bibitem{SPDC2}
C. W. S. Chang, C. Sabín, P. Forn-Díaz, F. Quijandría, A. M. Vadiraj, I. Nsanzineza, G. Johansson, C. M. Wilson, 
\textit{Phys. Rev. X} \textbf{2020}, 10, 011011. 

\bibitem{SPDC3}
G. Buser, R. Mottola, B. Cotting, J. Wolters, and P. Treutlein,
\textit{PRX Quantum} \textbf{2022}, 3, 020349.

\bibitem{SPDC4}
T. Santiago-Cruz, S. D. Gennaro, O. Mitrofanov, S. Addamane, J. Reno, I. Brener, M. V. Chekhova, 
\textit{Science} \textbf{2022}, 377, 991.

\bibitem{SPDC5}
H.-C. Weng, C.-S. Chuu,
\textit{Phys. Rev. Appl.} \textbf{2024}, 22, 034003.


\bibitem{SFWM1}
C.-Y. Hsu, Y.-S. Wang, J.-M. Chen, F.-C. Huang, Y.-T. Ke, E. K. Huang, W. Hung, K.-L. Chao, S.-S. Hsiao, Y.-H. Chen, C.-S. Chuu, Y.-C. Chen, Y.-F. Chen, I. A. Yu,
\textit{Opt. Express} \textbf{2021}, 29, 4632.

\bibitem{SFWM2}
J.-M. Chen, C.-Y. Hsu, W.-K. Huang, S.-S. Hsiao, F.-C. Huang, Y.-H. Chen, C.-S. Chuu, Y.-C. Chen, Y.-F. Chen, I. A. Yu,
\textit{Phys. Rev. Res.} \textbf{2022}, 4, 023132.

\bibitem{SFWM3}
S.-S. Hsiao, W.-K. Huang, Y.-M. Lin, J.-M. Chen, C.-Y. Hsu, I. A. Yu,
\textit{Phys. Rev. A} \textbf{2022}, 106, 023709.

\bibitem{SFWM4}
H. Jeong, H. Kim, H. S. Moon,
\textit{Adv. Quantum Technol.} \textbf{2023}, 7, 2300108.

\bibitem{SFWM5}
O. Davidson, O. Yogev, E. Poem, O. Firstenberg,
\textit{Phys. Rev. Lett.} \textbf{2023}, 131, 033601.

\bibitem{SFWM6}
A. N. Craddock, Y. Wang, F. Giraldo, R. Sekelsky, M. Flament, M. Namazi,
\textit{Phys. Rev. Appl.} \textbf{2024}, 21, 034012.

\bibitem{SFWM7}
Y.-S. Wang, K.-B. Li, C.-F. Chang, T.-W. Lin, J.-Q. Li, S.-S. Hsiao, J.-M. Chen, Y.-H. Lai, Y.-C. Chen, Y.-F. Chen, C.-S. Chuu, I. A. Yu, 
\textit{APL Photonics} \textbf{2022}, 7, 126102. 


\bibitem{app1} 
S. Zhou, S. Zhang, C. Liu, J. F. Chen, J. Wen, M. M. T. Loy, G. K. L. Wong, S. Du, 
\textit{Opt. Express} \textbf{2012}, 20, 24124. 

\bibitem{app2}
Z.-Y. Liu, Y.-H. Chen, Y.-C. Chen, H.-Y. Lo, P.-J. Tsai, I. A. Yu, Y.-C. Chen, Y.-F. Chen,
\textit{Phys. Rev. Lett.} \textbf{2016}, 117, 203601. 

\bibitem{app3}
Y.-F. Hsiao, P.-J. Tsai, H.-S. Chen, S.-X. Lin, C.-C. Hung, C.-H. Lee, Y.-H. Chen, Y.-F. Chen, I. A. Yu, Y.-C. Chen, 
\textit{Phys. Rev. Lett.} \textbf{2018}, 120, 183602. 

\bibitem{app4}
C.-Y. Cheng, J.-J. Lee, Z.-Y. Liu, J.-S. Shiu, Y.-F. Chen,
\textit{Phys. Rev. A} \textbf{2021}, 103, 023711. 

\bibitem{app5}  
C.-Y. Cheng, Z.-Y. Liu, P.-S. Hu, T.-N. Wang, C.-Y. Chien, J.-K. Lin, J.-Y. Juo, J.-S. Shiu, I. A. Yu, Y.-C. Chen, Y.-F. Chen,
\textit{Opt. Lett.} \textbf{2021}, 46, 681. 

\bibitem{app6}
K.-F. Chang, T.-P. Wang, C.-Y. Chen, Y.-H. Chen, Y.-S. Wang, Y.-F. Chen, Y.-C. Chen, I. A. Yu,
\textit{Phys. Rev. Res.} \textbf{2021}, 3, 013096.

\bibitem{app7}
Z.-Y. Liu, J.-S. Shiu, C.-Y. Cheng, Y.-F. Chen,
\textit{Phys. Rev. A} \textbf{2023}, 108, 013702.

\bibitem{app8}
P.-H. Tseng, L.-C. Chen, J.-S. Shiu, Y.-F. Chen,
\textit{Phys. Rev. A} \textbf{2024}, 109, 043716. 

\bibitem{app9}
D.-G. Im, Y. Kim, Y.-H. Kim,
\textit{Opt. Express} \textbf{2021}, 29, 2348.

\bibitem{app10}
D. Kim, J. Park, C. Baek, S. K. Lee, H. S. Moon,
\textit{Optica Quantum} \textbf{2024}, 2, 288.

\bibitem{SR1}
T. Chaneli\`{e}re, D. N. Matsukevich, S. D. Jenkins, T. A. B. Kennedy, M. S. Chapman, A. Kuzmich, 
\textit{Phys. Rev. Lett.} \textbf{2006},  96, 093604. 

\bibitem{SR2}
R. T. Willis, F. E. Becerra, L. A. Orozco, S. L. Rolston,
\textit{Phys. Rev. A} \textbf{2010}, 82, 053842.

\bibitem{SR3}
B. Srivathsan, G. K. Gulati, B. Chng, G. Maslennikov, D. Matsukevich, C. Kurtsiefer,
\textit{Phys. Rev. Lett.} \textbf{2013}, 111, 123602.

\bibitem{SR4}
Y.-S. Lee, S. M. Lee, H. Kim, H. S. Moon,
\textit{Phys. Rev. A} \textbf{2017}, 96, 063832.


\bibitem{osci1}
P. Kolchin, S. Du, C. Belthangady, G. Y. Yin, S. E. Harris,
\textit{Phys. Rev. Lett.} \textbf{2006}, 97, 113602.

\bibitem{osci2}
S. Du, J. Wen, M. H. Rubin, G. Y. Yin,
\textit{Phys. Rev. Lett.} \textbf{2007}, 98, 053601.

\bibitem{osci3}
M. O. Ara\'{u}jo, L. S. Marinho, D. Felinto,
\textit{Phys. Rev. Lett.} \textbf{2022}, 128, 083601.


\bibitem{pre1}
S. Du, P. Kolchin, C. Belthangady, G. Y. Yin, S. E. Harris,
\textit{Phys. Rev. Lett.} \textbf{2008}, 100, 183603.

\bibitem{pre2}
S. Zhang, J. F. Chen, C. Liu, M. M. T. Loy, G. K. L. Wong, S. Du,
\textit{Phys. Rev. Lett.} \textbf{2011}, 106, 243602.

\bibitem{pre3}
C. Liu, Y. Sun, L. Zhao, S. Zhang, M. M. T. Loy, S. Du,
\textit{Phys. Rev. Lett.} \textbf{2014}, 113, 133601.



\bibitem{Shiu1}
J.-S. Shiu, Z.-Y. Liu, C.-Y. Cheng, Y.-C. Huang, I. A. Yu, Y.-C. Chen, C.-S. Chuu, C.-M. Li, S.-Y. Wang, Y.-F. Chen,
\textit{Phys. Rev. Res.} \textbf{2024}, 6, L032001.

\bibitem{Shiu2}
J.-S. Shiu, C.-W. Lin, Y.-C. Huang, M.-J. Lin, I.-C. Huang, T.-H. Wu, P.-C. Kaun, Y.-F. Chen,
\textit{Phys. Rev. A} \textbf{2024}, 110, 063723.

\bibitem{SFWM8}
V. Bali\'{c}, D. A. Braje, P. Kolchin, G. Y. Yin, S. E. Harris, 
\textit{Phys. Rev. Lett.} \textbf{2005}, 94, 183601. 

\bibitem{SFWM9}
J. K. Thompson, J. Simon, H. Loh, V. Vuleti\'{c}, 
\textit{Science} \textbf{2006}, 313, 74. 

\bibitem{SFWM10}
J. Wen, M. H. Rubin, 
\textit{Phys. Rev. A} \textbf{2006}, 74, 023808. 

\bibitem{Kolchin}
P. Kolchin,
\textit{Phys. Rev. A} \textbf{2007}, 75, 033814.

\bibitem{SFWM11}
C. H. Raymond Ooi, Q. Sun, M. S. Zubairy, M. O. Scully, 
\textit{Phys. Rev. A} \textbf{2007}, 75, 013820. 

\bibitem{SFWM12}
Z. Han, P. Qian, L. Zhou, J. F. Chen, W. Zhang, 
\textit{Sci. Rep.} \textbf{2015}, 5, 9126. 

\bibitem{SFWM13}
A. Bruns, C.-Y. Hsu, S. Stryzhenko, E. Giese, L. P. Yatsenko, I. A. Yu, T. Halfmann, T. Peters,
\textit{Quantum Sci. Technol.} \textbf{2023}, 8, 015002.

\bibitem{SFWM14}
J.-M. Chen, T. Peters, P.-H. Hsieh, I. A. Yu,
\textit{Adv. Quantum Technol.} \textbf{2024}, 7, 2400138.

\bibitem{SFWM15}
W.-K. Huang, B. Kim, T.-J. Shih, C.-Y. Hsu, P.-Y. Tu, T.-Y. Lin, Y.-F. Chen, C.-S. Chuu, I. A. Yu,
\textit{Quantum Sci. Technol.} \textbf{2025}, 10, 015062.


\bibitem{EIT1}
S. E. Harris, J. E. Field, A. Imamoğlu,
\textit{Phys. Rev. Lett.} \textbf{1990}, 64, 1107.

\bibitem{EIT2}
M. Fleischhauer, A. Imamoglu, J. P. Marangos,
\textit{Rev. Mod. Phys.} \textbf{2005}, 77, 633.

\bibitem{EIT3}
H. Hsu, C.-Y. Cheng, J.-S. Shiu, L.-C. Chen, Y.-F. Chen,
\textit{Opt. Express} \textbf{2022}, 30, 2097.


\bibitem{Scully}
M. O. Scully, M. S. Zubairy,
\textit{Quantum Optics}, Cambridge University Press, Cambridge, \textbf{1997}.


\bibitem{Wick}
M. Gaudin, 
\textit{Nucl. Phys.} \textbf{1960} 15, 89.




\bibitem{SFWM16}
T.-J. Shih, W.-K. Huang, Y.-M. Lin, K.-B. Li, C.-Y. Hsu, J.-M. Chen, P.-Y. Tu, T. Peters, Y.-F. Chen, I. A. Yu,
\textit{Opt. Express} \textbf{2024}, 32, 13657.

\bibitem{SFWM17}
K.-S. Cui, X.-J. Zhang, and J.-H. Wu,
\textit{Phys. Rev. A} \textbf{2024}, 109, 063701.

\end{thebibliography}
\end{document}